\documentclass[iop]{emulateapj}
\shortauthors{Sekanina}
\shorttitle{Brightness and Orbital Motion of Comet C/2012 S1 (ISON)}
\slugcomment{Version October 7, 2013}
\newcommand{\Rsun}{$R_{\mbox{\scriptsize \boldmath $\odot$}}$}

\begin{document}
\title{Brightness and Orbital Motion Peculiarities of Comet C/2012 S1 (ISON):\\
 Comparison with Two Very Different Comets}
\author{Zdenek Sekanina}
\affil{Jet Propulsion Laboratory, California Institute of Technology,
4800 Oak Grove Drive, Pasadena, CA 91109, U.S.A.}
\email{Zdenek.Sekanina@jpl.nasa.gov}

\begin{abstract}
\noindent
To gain a greater insight into the impending evolution of the physical
behavior of comet C/2012~S1, its light curve and orbital properties
are compared with those for C/1962~C1 (Seki-Lines) and C/2002~O4
(H\"{o}nig).  All three are likely Oort Cloud comets.  C/1962~C1
survived an encounter with the Sun at less than 7~{\Rsun}, while
C/2002~O4 disintegrated near perihelion at 0.78 AU from the Sun.  Less
than two months before its perihelion at 2.7~{\Rsun}, C/2012~S1 has a
light curve that is much closer to C/1962~C1 than C/2002~O4.  It remains
to be seen whether its motion is affected by nongravitational perturbations.
As new data on C/2012~S1 keep coming in, its continuing comparison with
the two comets will provide information on its health by updating and
adjusting its status.  Strengths and weaknesses of this approach for
potential future applications to other comets will eventually be assessed.
\end{abstract}

\keywords{comets: general --- comets: individual (C/1953 X1, C/1962 C1,
C/1999 S4, C/2002 O4, C/2012 S1) --- methods: data analysis}

\section{Introduction: Choice of Comparison Comets}
Widespread concerns about the survival of comet C/2012~S1 during its
forthcoming close encounter with the Sun imply that every effort should
be expended to monitor the comet's physical behavior during its journey
to perihelion.  One way to contribute to this campaign is to compare,
step by step, the light curve of C/2012 S1 in the course of this time
with the light curves of other comets with very different histories, yet
of the same or similar origin. 

The first arrival from the Oort Cloud is a trait that C/2012~S1 shares with
C/1962~C1 (Seki-Lines) and probably also with C/2002~O4 (H\"{o}nig), which
makes these objects intriguing candidates for such comparison.  Besides the
light curve, there are also issues linked to the orbital motion of C/2012~S1
that include possible peculiarities and the perihelion distance of
0.0124~AU or 2.67~{\Rsun}.

Comet C/1962~C1 moved in an orbit with a perihelion distance closer to that of
C/2012~S1 than any other Oort cloud comet, merely 0.0314~AU or 6.75~{\Rsun},
and was thus subjected to thermal and radiation conditions almost --- though
not quite --- as harsh as are going to be experienced by C/2012~S1.  It
survived the encounter with its physique apparently intact.  On the other
hand, comet C/2002~O4 became famous (or, rather, infamous) by disappearing
(and obviously disintegrating) before the eyes of the observers almost exactly
at perihelion at 0.776~AU from the Sun..  In terms of approach to the Sun, a
better choice would have been C/1953~X1 (Pajdu\v{s}\'akov\'a), a disintegrating
comet whose perihelion distance was only 0.072~AU or 15.5~{\Rsun}.
Unfortunately, only a parabolic orbit is available for this object and the
Oort Cloud as the site of its origin is highly questionable.  In addition,
its light curve is, unlike that for C/2002~O4, only poorly known, appearing
rather flat over a period of at least 30~days and possibly as long as 70~days
(Sekanina 1984).  In fact, because of their early disappearance, the
disintegrating long-period comets have generally poor orbits.
% are singularly unsuitable for a quality orbit determination.
%  A disintegrating comet with a well determined orbit,
%  C/1999~S4 (LINEAR), was hardly an Oort Cloud object (Marsden 2000).
For the disintegrating comet C/1999 S4 (LINEAR), an exception to this rule,
use of the nongravitational terms in the equations of motion (Marsden 2000)
renders its original orbit indeterminate (Marsden et al.\ 1973).

In summary, C/1962 C1 and C/2002 O4 are the best available representatives of
two very different, almost extreme, categories of probable Oort Cloud comets
that I am aware of.  One disappointment with both comparison objects is that
they were discovered relatively late:\ C/1962~C1 only 56~days and C/2002~O4
just 72~days before perihelion.

A sequence of steps pursued in this investigation of C/2012~S1 begins in early
October 2013 with charting the outline, describing the primary objectives,
and addressing the specific issues examined.  This is the contents of the
paper itself.  This first step will be followed by a series of brief
contributions, to be appended, at a rate of about one per week until
mid-November (two weeks before the comet reaches perihelion), to the paper
as successive {\it Status Update Reports\/} based on newly available
information.  No predictions will be attempted, but systematic trends,
suggested by the most recent observations, will be pointed out.  While this
work is limited to only very particular tasks and is not intended to solve
the issue of the comet's survival, it should offer an opportunity to
potencially refocus the comet's monitoring programs and to gradually adjust
the prospects for survival chances.  Eventually --- at some point after
the show is over --- it will be useful to assess strengths and weaknesses
of this approach for improvements in potential future applications to other
exceptional comets.

\begin{figure*}
\vspace{-2.55cm}  % -3.15
\hspace{-0.35cm}  % -0.38
\centerline{
\scalebox{0.67}{ % 0.825
\includegraphics{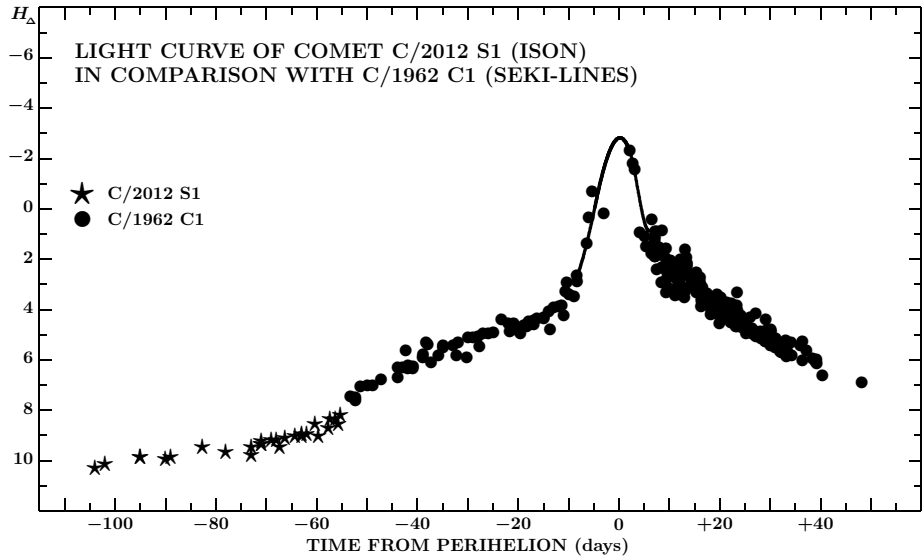}}} % from net_2012S1.tex
\vspace{-9.9cm}  % -12.2
\caption{Light curves of comets C/2012 S1 and C/1962 C1.  The total visual
magnitude normalized to 1~AU from the earth, $H_\Delta$, is plotted against
time from perihelion.  The perihelion times are November 28, 2013 for
C/2012~S1 and April~1, 1962 for C/1962~C1.  The brightness data for the two
objects are plotted with different symbols, as indicated.  The brightness
enhancement of C/1962~C1 around perihelion was due in part to forward
scattering of light by microscopic dust parlicles in the coma.{\vspace{0.4cm}}}

\end{figure*}

\section{The Light Curves}
A light curve is in this investigation understood to be a plot of a total
brightness (expressed in magnitudes), normalized to a distance $\Delta$ of
1~AU from the earth by employing the usual term \mbox{$5 \log \Delta$}, against
time or heliocentric distance.  The phase effect is not accounted for, but
its potential implications for the light curve are always described in the
text.  The magnitudes are referred to a visual spectral region.  Because
every observer measures a comet's brightness in his own photometric system,
this heterogeneity has to be eliminated (or reduced as much as possible)
by introducing corrections to a standard photometric system.  Its zero
point is defined by M.~Beyer's brightness data (see below), whose scale
is tied to the {\it International Photovisual System \/}(Ipv; e.g.,
Seares 1922).  His light curves of comets were employed not only to
calibrate the personal photometric systems of other observers of those
same comets (separately for each instrument used) but also extended ---
generally by observer/instrument-to-observer/instrument multiple chain
comparisons of overlapping rows of brightness estimates --- to fainter
magnitudes reported by observers (whether visual or those using CCD
detectors) of other comets, including the ones studied in this paper.
By multiply crosschecking time overlaps by the same observers it has
been possible, by introducing personal/instrumental corrections, to
largely remove major systematic magnitude differences for at least
some individuals, who then make up a check list of {\it pivotal\/}
observers,\footnote{For example, the total magnitudes that K.\ Kadota,
one of many involved in monitoring the motion and brightness of C/2012~S1,
reports from his CCD observations with a 25-cm f/5 reflector are fainter
than the scale of the adopted photometric system and require for comets
of a moderately condensed appearance a correction near $-$1.0~magnitude.}
to whose photometric scale the magnitude observations by others are
readily linked.  Compared to Beyer, most observers underestimate the
total brightness.  Corrections greater than $\sim$2~magnitudes are,
however, suspect and such data should not generally be employed in
light-curve analyses.

\section{Comet C/2012 S1 (ISON)}
To compare comet C/2012 S1 with the two other objects, there is no point in
presenting its early light curve, since the reference comets were discovered
less than 1.6~AU from the Sun.  Besides, the behavior of C/2012 S1 beyond 3~AU
was recently investigated by Ferrin (2013).

In this paper, the light curve is examined from the time after the comet
reemerged in August 2013 from its conjunction with the Sun.  A total of
28~magnitude observations, mostly from CCD images, was normalized in accord
with the procedure of Sec.~2.  They were made by five observers between
August~16 and October~4, or 104 to 55~days before the comet's perihelion
on November 28.  The observations were difficult, especially before the
end of August, when the comet was less than 32$^\circ$ from the Sun.  The
phase angle stayed between 9$^\circ$ and 29$^\circ$ during the 50~days of
observation, having no major effect on the light curve presented in Figure~1.

\section{Comet C/1962 C1 (= 1962 III = 1962{\tiny \bf c})}
Discovered independently by R.\ D.\ Lines and T.\ Seki within 8~hours of
each other, this comet was well observed right from the beginning.  Its
perihelion passage occurred on April 1, 1962.  The light curve,
presented in Figure~1, is based on 244 brightness estimates by 34~observers,
including Beyer (1963).  The phase angle was steadily increasing from
34$^\circ$ at the time of the first used brightness observation, on February~7,
to 90$^\circ$ some 17~days before perihelion, when the comet was 0.70~AU
from the Sun.  The Henyey-Greenstein law modified and applied to cometary
dust by Marcus (2007) suggests that if all light was due to dust, the phase
correction would have been between $-$0.5 and $-$0.9 magnitude in this period
of time.  Forward scattering then began to take over.  The phase angle reached
100$^\circ$ about 11~days before perihelion, 110$^\circ$ six days later, and
peaked near 115$^\circ$ at \mbox{2--3}~days before perihelion.  This is the
section of the light curve, where the rate of brightness increase
accelerated sharply.  The phase correction at the peak phase angle is about
+0.3~magnitude.  If these numbers are valid, the corrected peak brightness
would be, relative to much of the preperihelion light curve, about
1~magnitude less prominent.  However, most of the near-perihelion data
points were obtained in daylight or near sunset, being highly inaccurate.
By the time the comet reached perihelion, the phase angle dropped to
$\sim$80$^\circ$ and continued to decrease gradually, attaining 26$^\circ$
at the time of Beyer's last observation 48~days after perihelion, on May 19.

\begin{table}[b]
\noindent
\vspace{0.05cm}
\begin{center}
{\footnotesize {\bf Table 1}\\[0.08cm]
{\sc Twilight and Daytime Observations of Comet C/1962 C1 by\\A.\,D.\,Thackeray
 at Radcliffe Observatory, Pretoria,\\Attempted on March 29--April 1, 1962,
 and\\Bortle's limiting daylight magnitudes.}\\[0.12cm]
\begin{tabular}{l@{\hspace{0.07cm}}r@{\hspace{0.55cm}}c@{\hspace{0.35cm}}c@{\hspace{0.15cm}}c@{\hspace{0.6cm}}c}
\hline\hline\\[-0.25cm]
\multicolumn{2}{@{\hspace{-0.2cm}}c}{Time of} & Time from
 & \multicolumn{2}{@{\hspace{-0.37cm}}c}{Magnitude (mag)} & \\[-0.03cm]
\multicolumn{2}{@{\hspace{-0.2cm}}c}{observation} & perihelion
 & \multicolumn{2}{@{\hspace{-0.37cm}}c}{\rule[0.6ex]{2.6cm}{0.4pt}}
 & Elongation \\[-0.02cm]
\multicolumn{2}{@{\hspace{-0.2cm}}c}{1962 (UT)} & (days) & observed$^{\rm a}$
 & limiting & from Sun \\[0.05cm]
\hline\\[-0.22cm]
March & 29.70 & $-$2.96 & \llap{$\sim$}0 & $-$0.4\rlap{$^{\rm b}$}
      & \llap{1}0$^\circ\!\!$.7 \\
      & 30.67 & $-$1.99 & U &  $-$0.7  & 7.9 \\
      & 31.5$\;\:$ & $-$1.16 &  U &  $-$1.1  & 5.3 \\
April &  1.5$\;\:$ & $-$0.16 &  U &  $-$1.9  & 2.0 \\[0.05cm]
\hline\\[-0.28cm]
\multicolumn{6}{l}{\parbox{7.5cm}{$^{\rm a}$\, \scriptsize U means the comet
 was undetected.}}\\[-0.08cm]
\multicolumn{6}{l}{\parbox{8.2cm}{$^{\rm b}$\, \scriptsize Not a daytime
observation; Bortle formula not strictly applicable.}}\\[-0.4cm]
\end{tabular}}
\end{center}
\end{table}

One of the observations plotted in Figure 1 was made by A.~D.~Thackeray at
the Radcliffe Observatory in Pretoria, South Africa, with an 8-cm refractor
on March 29.70 UT, 71 hours before perihelion and only 43~minutes after
sunset.  The comet was 10$^\circ\!$.7 from the Sun and the observation, made
through a rift in the clouds, was reported by Venter (1962).  He remarked
that \mbox{Thackeray} also attempted, to detect the comet with an 18-cm finder
of the 188-cm reflector of the observatory at sunset on March~30 and with
binoculars in broad daylight on March~31 and April~1, always unsuccessfully.
These observations, positive or negative alike, are interesting to compare
with Bortle's (1985) efforts to determine a limiting magnitude $H_{\rm lim}$
for the faintest cometary objects detectable in daylight as a function of
the elongation from the Sun (see also Green 1997).  While Bortle's test
observations were made on Mercury and Venus, he emphasized that when very
slightly defocused, the two planets rather closely mimic the appearance of
a daylight comet.  He concluded that although his experiments were conducted
with 8-cm binoculars, the predicted relationship is not strongly aperture
dependent because of the brightness of the sky background.  However, the
practical result does depend on how well the Sun is occulted during the
observation and especially how the instrument's optical surfaces are
protected against direct illumination.

\begin{table}[b]
\vspace{0.3cm}
\noindent
\begin{center}
{\footnotesize {\bf Table 2}\\[0.08cm]
{\sc Time from Perihelion and Heliocentric Distance\\for Orbits of Three
 Investigated Comets.}\\[0.10cm]
\begin{tabular}{c@{\hspace{0.85cm}}c@{\hspace{0.85cm}}c@{\hspace{0.85cm}}c}
\hline\hline\\[-0.22cm]
Time from 
 & \multicolumn{3}{@{\hspace{-0.02cm}}c}{Heliocentric distance (AU)}\\[-0.03cm]
perihelion
 & \multicolumn{3}{@{\hspace{-0.02cm}}c}{\rule[0.6ex]{6.1cm}{0.4pt}}\\[-0.02cm]
(days) & C/2012 S1 & C/1962 C1 & C/2002 O4\\[0.05cm]
\hline\\[-0.22cm]
$\;\:$0    & 0.012 & 0.031 & 0.776 \\
10         & 0.498 & 0.481 & 0.800 \\
20         & 0.798 & 0.780 & 0.867 \\
30         & 1.050 & 1.032 & 0.965 \\
40         & 1.274 & 1.256 & 1.083 \\
50         & 1.481 & 1.462 & 1.212 \\
60         & 1.674 & 1.655 & 1.347 \\
70         & 1.856 & 1.838 & 1.484 \\
80         & 2.030 & 2.012 & 1.622 \\
90         & 2.197 & 2.179 & 1.760 \\
\llap{1}00 & 2.358 & 2.339 & 1.896 \\
\llap{1}10 & 2.513 & 2.495 & 2.030 \\[0.08cm]
\hline\\[-0.65cm]
\end{tabular}}
\end{center}
\end{table}

The four instances of Thackeray's effort to detect the comet in daylight or
twilight are listed in Table~1.  The times of his last two observations are
not known and are assumed to be early in the afternoon local time.  The
successful observation on March~29 does not satisfy the conditions for
applying Bortle's formula, nevertheless each of the comparisons provides
useful information.

We do not know how bright in fact the comet was around perihelion, but the
curve fitted through the nearby points in Figure~1 suggests for Thackeray's
search times the magnitudes $-$2.2 for March~30, $-$2.5 for March~31, and
$-$2.8 for April~1, that is, the comet brighter than the limiting magnitudes
by, respectively, 1.5, 1.4, and 0.9~magnitudes.  However, Thackeray's March 29
estimate is 1.6 magnitudes much too faint (as seen from Figure~1), so the
disagreement is not surprising.  Either Thackeray did not carry out his
daytime observations of C/1962~C1 in compliance with Bortle's conditions or
the comet was in close proximity to the Sun{\nopagebreak} fainter than the
light curve in Figure~1 indicates.

\section{Comet C/2002 O4 (H\"{o}nig)}
The results for C/2002 O4 are taken from my earlier paper (Sekanina 2002), to
which the reader is referred in order to learn more about the idiosyncrasies
of this object.  The phase angle stayed during the observations between
39$^\circ$ and 68$^\circ$, with no major effect on the shape of the light
curve, which, however, is not plotted in Figure~1 because the comparison with
the other two objects would be meaningless.  Since activity of a comet depends
on the Sun's radiation input to the nucleus, the light curves for comets of
very different perihelion distances need to be compared against heliocentric
distance, not time.  This is clearly seen from Table~2, where the relationship
between time and heliocentric distance is shown for the three investigated
comets.  While this relationship is nearly identical for C/2012~S1 and
C/1962~C1 because of their similar perihelion distances, the situation is
very different for C/2002~O4, whose scale of heliocentric distances is
strongly compressed.   The comet needs 30 more days than C/2012~S1 to get
from 2~AU to perihelion.

\begin{figure*}
\vspace{-2.6cm} % -3.15
\hspace{-0.4cm}
\centerline{
\scalebox{0.67}{ % 0.825
\includegraphics{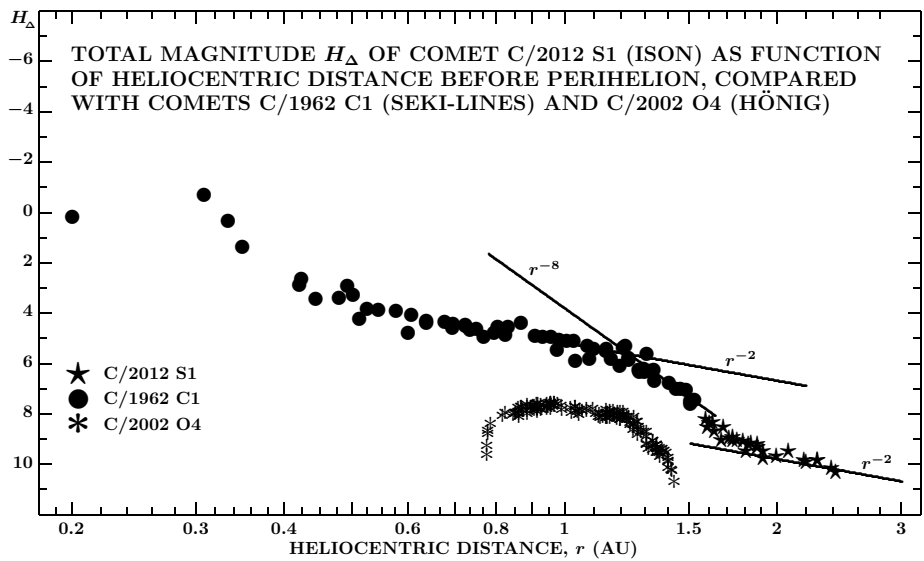}}} % from net_2012S1r.tex
\vspace{-9.9cm} % -12.2
\caption{Plot of preperihelion light curves of comets C/2012 S1, C/1962~C1,
and C/2002~O4 against heliocentric distance $r$.  The total visual magnitude,
$H_\Delta$, is again normalized to 1\,AU from the earth.  The perihelion
distances of the three comets are, respectively, 0.012\,AU or 2.7\,{\Rsun},
0.031\,AU or 6.7\,{\Rsun}, and 0.78\,AU.  Their brightness data are plotted
with different {\vspace{-0.06cm}}symbols, as indicated.  Also depicted are
slopes of brightness variations proportional to $r^{-2}$ and $r^{-8}$.  The
upswing at \mbox{$r < 0.5$\,AU} on the light curve of C/1062~C1 is due in
part to forward scattering (at phase angles $>100^\circ$) of light by
microscopic dust particles in the coma.{\vspace{0.4cm}}}
\end{figure*}

\section{Brightness Variation with Distance from Sun}
Figure 2 compares the preperihelion light curves of the three comets, with
the heliocentric distance being plotted instead of time on the axis of
abscissae.  For comet C/2002~O4 the discovery magnitude, whose photometric
correction is practically impossible to determine, has been omitted in the
plot.

The figure displays two major differences between the light curves of
C/1962~C1 and C/2002~O4 (which happened to be one of the brighter
disintegrating comets).  The first difference is in the brightness:\ at
any heliocentric distance, C/1962~C1 was intrinsically brighter by at least
2~magnitudes.  The second difference is in the shape of the light curve:\
the brightness of C/1962 C1 kept continuously climbing with its approach
to the Sun, while that of C/2002~O4 began to stall about one month before
perihelion, when the comet was at a heliocentric distance of about 0.96~AU,
bending eventually downward at an accelerating rate.

The relation between C/2012~S1 and C/1962~C1 is very similar to that in
Figure~1.  The forthcoming magnitude data will determine whether the light
curve of C/2012~S1 will essentially coincide with that of C/1962~C1 or
will extend below or above it.  The data will also show whether the two
light curves will or will not be nearly parallel.  Most importantly, the
upcoming observations should reveal any possible tendency toward brightness
stalling, which would be a sign of disappointing performance near the Sun.
By early October, no such effect is obvious from a nearly two-month arc
covered by the post-conjunction data set.  It is noticed from Figure~2 that
the rate of brightening with decreasing heliocentric distance is a little
steeper than $r^{-2}$, an encouraging indicator.

Another gratifying sign is that comet C/2012~S1 is intrinsically much
brighter at 1.6~AU from the Sun than C/2002~O4 was at 1.4~AU.  And if the
early part of the light curve of this latter object was a result of an
outburst (Sekanina 2002), then C/2012~S1 is {\it considerably\/} brighter
than was C/2002~O4 at the same distance.

\section{Nongravitational Effects in Orbital Motion}
It is true, though odd, that the light curve of comet C/2002~O4 covers
a time interval two weeks longer than the arc covered by the astrometric
observations.  This is partly because the astrometry started only five days
after discovery, but mainly because the comet's brightness could still be
estimated after its nuclear condensation disappeared and there was nothing
to bisect for the position.  The loss of the condensation is the most
ominous attribute of disintegrating comets.  Unfortunately, this
unmistakable sign of imminent demise sets in rather suddenly and near the
end of a rapidly progressing process, so it is definitely not an early
warning sign.

\begin{table}[b]
\noindent
\begin{center}
{\footnotesize {\bf Table 3}\\[0.08cm]
{\sc Original Reciprocal Semimajor Axis for Comet C/2002 O4\\As Function
 of Orbital Solution's End Date.}\\[0.10cm]
\begin{tabular}{l@{\hspace{0.12cm}}c@{\hspace{0cm}}r@{\hspace{1.23cm}}c@{\hspace{0.93cm}}c}
\hline\hline\\[-0.22cm]
\multicolumn{3}{@{\hspace{-0.9cm}}c}{End date} & Original reciprocal
 & Number \\[-0.03cm]
\multicolumn{3}{@{\hspace{-0.9cm}}c}{of orbital} & semimajor axis,
 & of observa- \\[-0.02cm]
\multicolumn{3}{@{\hspace{-0.9cm}}c}{solution} & $(1/a)_{\rm orig}$\,(AU$^{-1}$)
 & tions used \\[0.05cm]
\hline\\[-0.22cm]
2002 & Sept. &  2 & $-$0.000\,520$\;\pm\;$0.000\,096 & 946 \\
     &       & 10 & $-$0.000\,694$\;\pm\;$0.000\,054 & 984 \\
     &       & 13 & $-$0.000\,717$\;\pm\;$0.000\,031 & \llap{1}088 \\
     &       & 23 & $-$0.000\,772$\;\pm\;$0.000\,021 & \llap{1}135 \\[0.08cm]
\hline\\[-0.6cm]
\end{tabular}}
\end{center}
\end{table}

A better timely indicator of the impending termination of a comet's existence
is a {\it major, progressively increasing\/} deviation of its orbital motion
from the gravitational law.  Although detection of these {\it nongravitational
perturbations\/} requires a fairly high-quality orbit determination, they
begin to show up much earlier than the condensation's disappearance.  To detect
this effect, there is no need to solve for the nongravitational parameters
(often a doomed effort); a straightforward approach is to look for temporal
variations in the reciprocal semimajor axis, $1/a$, derived from different
orbital arcs.  A {\it clear\/} systematic trend toward smaller $1/a$ with
time, as the end date of the observations included in a set of gravitational
orbital solutions is stepwise advanced, is a sign that the comet is in
trouble.  This trend means that the comet orbits the Sun in a gravity field
of decreasing magnitude, the deviations apparently caused by a momentum
transferred to the eroding nucleus by sublimation- and fragmentation-driven
forces, in the final stage probably also by solar radiation pressure.
Negative values of the original reciprocal semimajor axis $(1/a)_{\rm orig}$
are particularly worrisome.  One has to be sure, however, that this is not
due to inaccurate data and that the magnitude of the effect clearly exceeds
the errors of observation.

Even though comet C/2002~O4 was under observation for only about two months,
Marsden (2002a, 2002b) successively derived four general orbital solutions.
Each of them used astrometric observations that covered an orbital arc
starting on July~27 but ending at different times.  The resulting values
of $(1/a)_{\rm orig}$ that Marsden obtained were summarized in my previous
work on the comet (Sekanina 2002) and are in abbreviated form presented in
Table~3.  The hyperbolic excess, driven by the nongravitational forces,
which was already enormous in the first solution, grew further by
$\sim$250~units of 10$^{-6}$\,AU$^{-1}$ as the end date of the subsequent
solutions
advanced by only three weeks.

\begin{table}[t]
\noindent
\vspace{-0.25cm}
\begin{center}
{\footnotesize {\bf Table 4}\\[0.08cm]
{\sc Original Reciprocal Semimajor Axis for Comet C/2012~S1\\As Function of
 Orbital Solution's End Date.}\\[0.10cm]
\begin{tabular}{l@{\hspace{0.08cm}}c@{\hspace{0.02cm}}r@{\hspace{0.33cm}}c@{\hspace{-0.04cm}}c@{\hspace{0.04cm}}l}
\hline\hline\\[-0.22cm]
\multicolumn{3}{@{\hspace{0cm}}c}{End date} & Original reciprocal
 & Number & \\[-0.03cm]
\multicolumn{3}{@{\hspace{0cm}}c}{of orbital} & semimajor axis,
 & of observa- & \\[-0.02cm]
\multicolumn{3}{@{\hspace{0cm}}c}{solution}
 & $(1/a)_{\rm orig}$\,(AU$^{-1}$) & tions used
 & $\;$Reference$^{\rm a}$\\[0.05cm]
\hline\\[-0.22cm]
2012 & Oct. & 25 & $+0.000\:056\,9\pm0.000\:016\,2$ & $\;\:$418 & MPC\,80809 \\
     & Dec. & 24 & $+0.000\:039\,4\pm0.000\:008\,8$ & 1000 & MPC\,81859 \\
2013 & Jan. & 24 & $+0.000\:045\,2\pm0.000\:005\,4$ & 1612 & MPC\,82319 \\
     & Feb. & 20 & $-0.000\:000\,6\pm0.000\:003\,6$ & 2372 & MPC\,82720 \\
     & Mar. & 23 & $-0.000\:012\,2\pm0.000\:002\,5$ & 3181 & MPC\,83144 \\
     & Apr. & 20 & $-0.000\:000\,7\pm0.000\:002\,0$ & 3442 & MPC\,83520 \\
     & June &  8 & $+0.000\:007\,1\pm0.000\:001\,6$ & 3722 & MPC\,84317 \\
     & Aug. & 18 & $+0.000\:008\,5\pm0.000\:001\,6$ & 3740 & MPC\,84625 \\
     & Aug. & 23\rlap{$^{\rm b}$} & $+0.000\:028\,6\pm0.000\:001\,4$ & 3746
     & E\,2013-Q27 \\
     & Sept. & 6\rlap{$^{\rm b}$} & $+0.000\:009\,2\pm0.000\:001\,2$ & 3897
     & E\,2013-R59 \\
     & Sept. & 16\rlap{$^{\rm b}$} & $+0.000\:008\,6\pm0.000\:001\,1$
     & 3997 & MPC\,84932 \\
     & Sept. & 30\rlap{$^{\rm b}$} & $+0.000\:005\,4\pm0.000\:000\,7$
     & 4308 & E\,2013-S75 \\
\hline\\[-0.22cm]
\multicolumn{6}{l}{\parbox{8.26cm}{$^{\rm a}$\,\scriptsize MPC = Minor Planet
 Circular; E = Minor Planet Electronic Circular (MPEC).}}\\[0.12cm]
\multicolumn{6}{l}{\parbox{8.2cm}{$^{\rm b}$\,\scriptsize Start date for this
 solution was 2011 Sept.\ 30.}}\\[0.3cm]
\end{tabular}}
\end{center}
\end{table}

While the nominal values of $(1/a)_{\rm orig}$ leave no doubt that comet
C/2002~O4 could not have its aphelion nearer than the Oort Cloud, an effort
aimed at interpreting the steep rate of change in $(1/a)_{\rm orig}$ in
Table~3 led the author to a somewhat more ambiguous conclusion (\mbox{Sekanina}
2002).  While the Oort Cloud origin was still the most likely, the
uncertainty of the result was much too high.  On the other hand, I am
unaware of a candidate better than C/2002~O4 for a disintegrating Oort
Cloud comet.

For comparison, the original reciprocal semimajor axis from a dozen orbital
solutions for C/2012~S1 is listed in Table~4, in which their start date
(that is, the first astrometric observation used) is December~28, 2011,
unless stated otherwise.  Overall, no clear trend is perceived.  However,
when one considers only the last four entries, which are based on an
orbital arc extended further back in time compared to the previous ones
after the comet's 11~images were identified on CCD frames taken at the
Haleakala Pan-STARRS Station from September~30, November~10 and 26, and
December~9, 2011 (see MPEC 2013-Q27), a systematic negative trend is
apparent in $(1/a)_{\rm orig}$ at an average rate of about 18 units of
10$^{-6}$\,(AU)$^{-1}$ per month of the end-date advance.  This is much
less than the rate for C/2002~O4 and most of it appears to have occurred
between the end dates of August~23 and September~6, but the accuracy of
this result is also much higher than for C/2002~O4.  The future develop\-ment
of this potentially unsettling issue needs to be monitored.

\section{Conclusions Based on Observations Up to Early October 2013}
Comparison of C/2012~S1 with two very different comets shows that, as
of early October, its intrinsic brightness is close to that of C/1962~C1,
another Oort Cloud comet, which survived perihelion at a distance
4~{\Rsun} greater than is C/2012~S1's.  The forthcoming weeks will show
whether one can become more confident about potential similarities between
these two objects.

On the other hand, indications are that C/2012~S1 will intrinsically be
significantly brighter in the coming several days than C/2002~O4 was at
the same heliocentric distance shortly after discovery (\mbox{$H_\Delta
= 10.7$}~AU; equivalent to October~12 for C/2012~S1).  On the other hand,
comet C/1999~S4, perhaps the most prominent disintegrating comet (other than
the disintegrating sungrazers), was between 2.4 and 2.2~AU from the Sun just
about as bright intrinsically as C/2012~S1.  However, the rate of brightening
of C/1999~S4 between 2.2 and 1.6~AU, where it was picked up after conjunction
with the Sun, was so sluggish that by 1.6~AU (around October~3) comet
C/2012~S1 was already significantly brighter.

Throughout the remaining preperihelion orbital arc of C/2012~S1, two issues
to pay close attention to --- among others that may come to the forefront of
interest --- are:\ (i)~its rate of brightening (or fading?), as mentioned
above, and (ii)~the exact nature of its orbital motion, especially in terms
of systematic changes in the original semimajor axis.  An unwelcome sign
would be the need to introduce the nongravitational terms into the equations
of motion in order to fit satisfactorily the newly available astrometric
observations.  If the slight tendency of sliding toward ever more negative
values continues or even accelerates, concerns about the prospects for an
impressive show near perihelion, will be warranted.

In approximately weekly intervals until mid-November, {\it Status Update
Reports\/} (SURs) will be appended to this paper, based on results from the
most recent relevant observations.  The SURs will include updates to
Figures~1 and 2 and to a truncated Table 4, which will include only the
orbital solutions with the start date of September 30, 2011.\\[-0.2cm]

This research was carried out at the Jet Propulsion Laboratory, California
Institute of Technology, under contract with the National Aeronautics and
Space Administration.\\[-0.2cm]

\begin{center}
{\footnotesize REFERENCES}
\end{center}
\vspace{0.1cm}
{\footnotesize
\parbox{8.6cm}{Beyer, M. (1963).  Physische Beobachtungen von Kometen. XIII.
{\hspace*{0.3cm}}{\it Astron. Nachr.\/} {\bf 287}, 153--167.}\\[0.04cm]
\parbox{8.6cm}{Bortle, J. E. (1985).  The Observation of Bodies in Close
 Proximity {\hspace*{0.3cm}}to the Sun.  {\it Int. Comet Quart.\/} {\bf 7},
 7--11.}\\[0.04cm]
\parbox{8.6cm}{Ferrin, I. (2013).  The Location of Oort Cloud Comets C/2011~L4
 {\hspace*{0.3cm}}Panstarrs and C/2012~S1 ISON, on a Comets' Evolutionary
 Di- {\hspace*{0.3cm}}agram. {\it eprint} {\tt arXiv:1306.5010.}}\\[0.04cm]
\parbox{8.6cm}{Green, D. W. E., ed. (1997).  Daytime Observations of Comets,
 in {\hspace*{0.3cm}}{\it The ICQ Guide to Observing Comets\/}, Smithsonian
 Astrophys- {\hspace*{0.3cm}}ical Observatory, Cambridge, Mass., pp.\
 90--92.}\\[0.04cm]
\parbox{8.6cm}{Marcus, J. N. (2007).  Forward-Scattering Enhancement of
 Comet {\hspace*{0.3cm}}Brightness.  I.\ Background and Model. {\it Int.
 Comet Quart.\/} {\bf 29}, {\hspace*{0.3cm}}39--66.}\\[0.04cm]
\parbox{8.6cm}{Marsden, B. G. (2000).  Comet C/1999 S4 (LINEAR).  {\it Minor
 Plan. {\hspace*{0.3cm}}Circ.\/} 40988.}\\[0.04cm]
\parbox{8.6cm}{Marsden, B. G. (2002a).   Comet C/2002 O4 (Hoenig), {\it
 Minor Plan. {\hspace*{0.3cm}}Electr. Circ.\/} 2002-R15, 2002-R48, and
 2002-S10.}\\[0.04cm]
\parbox{8.6cm}{Marsden, B. G. (2002b).  Comet C/2002 O4 (Hoenig), {\it
 Minor Plan. {\hspace*{0.3cm}}Circ.\/} 46762.}\\[0.04cm]
\parbox{8.6cm}{Marsden, B. G.; Z. Sekanina; and D. K. Yeomans (1973).
 Comets {\hspace*{0.3cm}}and Nongravitational Forces.\ V, {\it Astron. J.\/}
 {\bf 78}, 211--225.}\\[0.04cm]
\parbox{8.6cm}{Seares, F. H. (1922).   Report of the President, Commission~25:
 {\hspace*{0.3cm}}Stellar Photometry. {\it Trans. Int. Astron. Union\/}
 {\bf 1}, 69--82.}\\[0.04cm]
\parbox{8.6cm}{Sekanina, Z. (1984).  Disappearance and Disintegration of
 Comets. {\hspace*{0.3cm}}{\it Icarus\/} {\bf 58}, 81--100.}\\[0.04cm]
\parbox{8.6cm}{Sekanina, Z. (2002).  What Happened to Comet C/2002~O4
 {\hspace*{0.3cm}}(H\"{o}nig)?  {\it Int. Comet Quart.\/} {\bf 24},
 223--236.}\\[0.04cm]
\parbox{8.6cm}{Venter, S. C. (1962).  Observations of Comet Seki-Lines 1962c.
 {\hspace*{0.3cm}}{\it Mon. Notes Astron. Soc. South Africa\/} {\bf 21},
 89--91.}\\[-0.2cm]
\clearpage
\mbox{ }\\[-0.79cm]
\begin{center}
{\large \bf STATUS UPDATE REPORT \#1}\\
{\large \bf (October 15, 2013)}\\
\end{center}
{\normalsize
This {\it SUR\/} appends results from the relevant observations of C/2012~S1
made between October 4, the cutoff date in the paper, and October~13.
Numerous reports of the comet's total magnitude measurements have been
made available, mostly via CCD imaging.   The number of observations
used in the updated light curve increased from 28 in the paper to 44 now,
based on the additional data by six observers.  The photometric-system
correction for one of the six has been refined.  The updated light curve
is presented in Figure~SUR1-1 in both versions, with the normalized
magnitude $H_\Delta$ plotted against time in the upper panel and against
heliocentric distance in the lower panel.  A few of the new data points
date before October~4.  The figure shows that the light curve of C/2012~S1
is currently running a little more than 1~magnitude below C/1962~C1.
% {\vspace{1cm}\pagebreak}

\begin{figure}[b]
\vspace{-1.4cm}
\hspace{-0.27cm}
\centerline{
\scalebox{0.475}{
\includegraphics{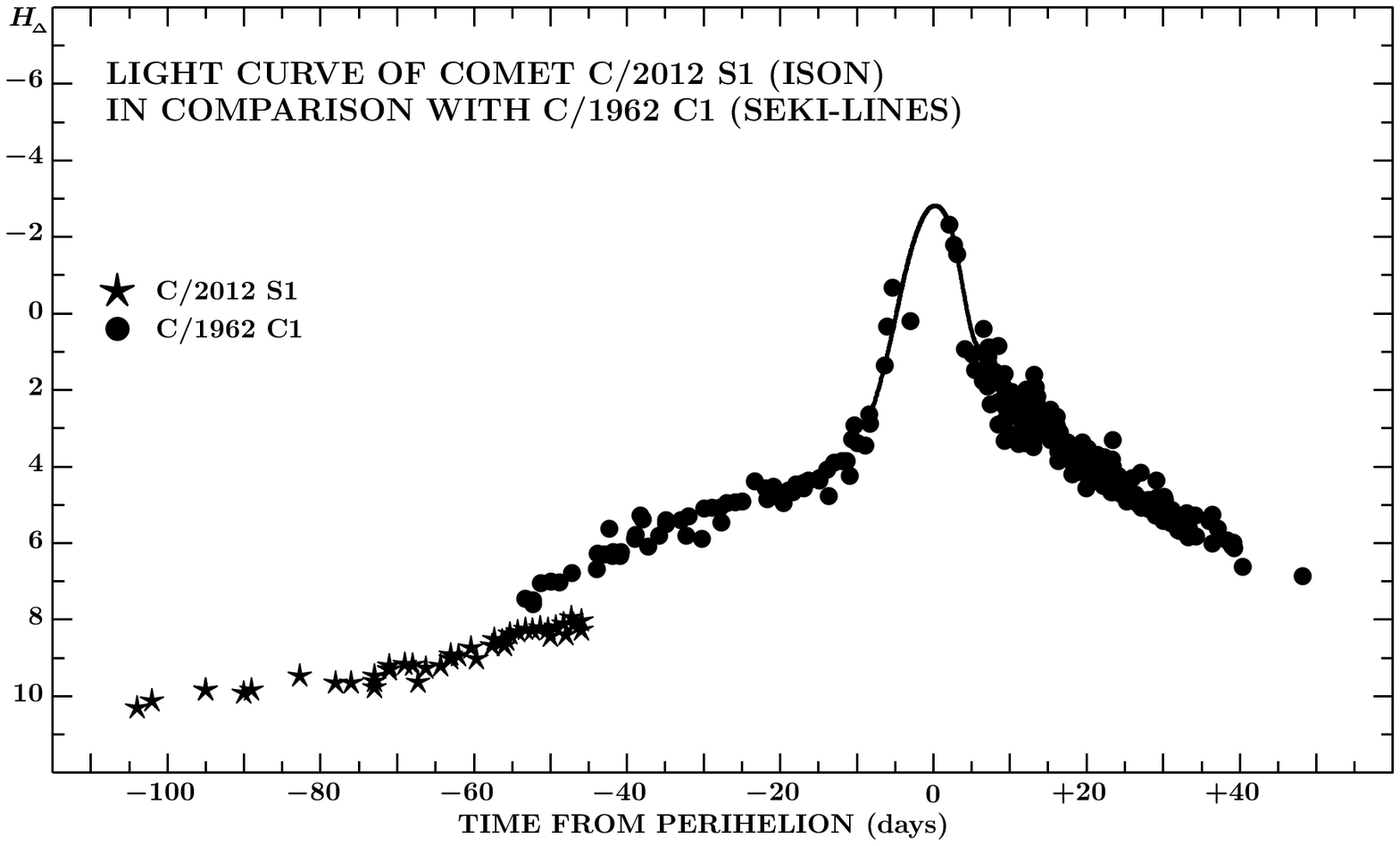}}} % from net_2012S1.tex

\vspace{-8.5cm}
\hspace{-0.305cm}
\centerline{
\scalebox{0.475}{
\includegraphics{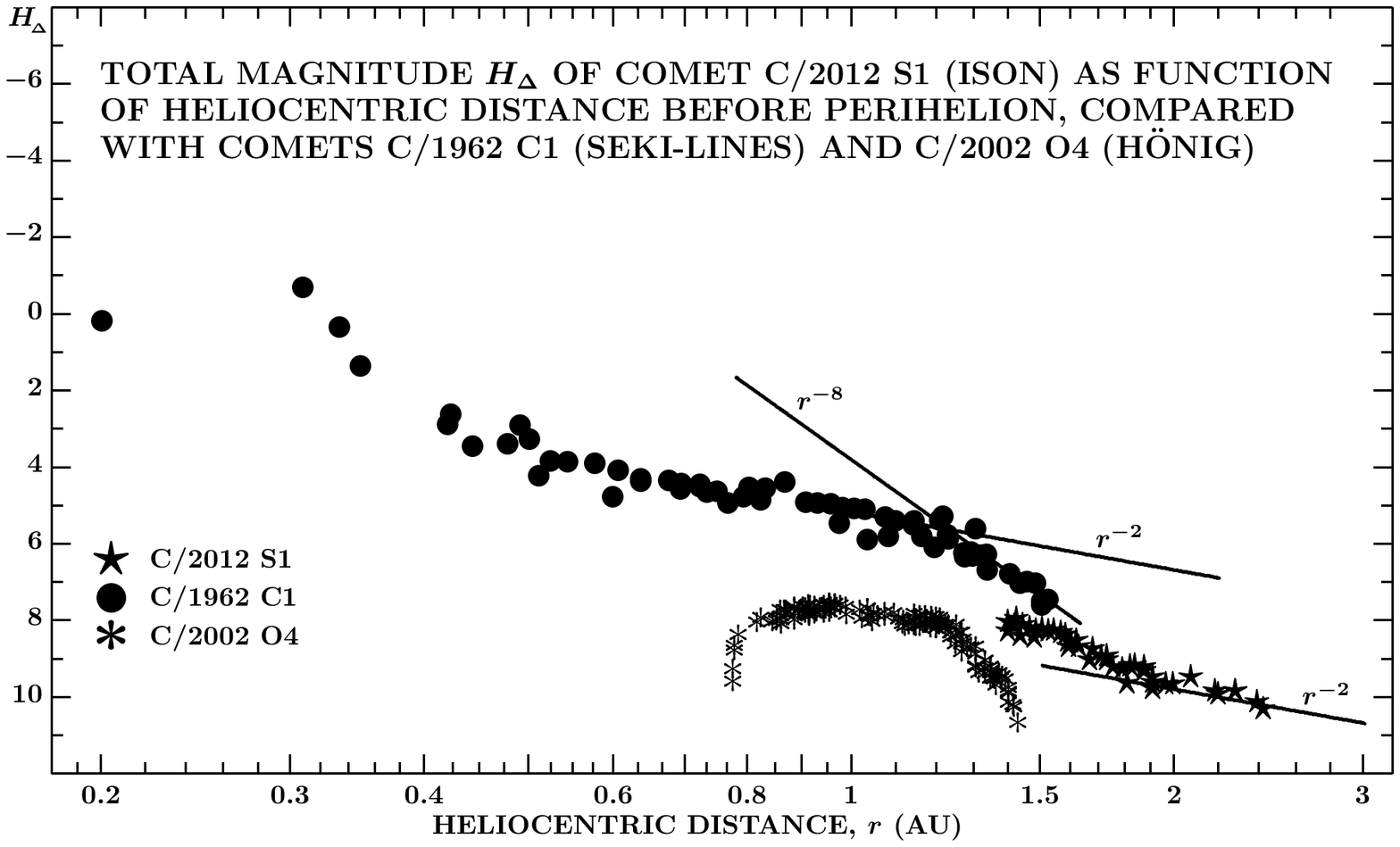}}} % from net_2012S1r.tex

\vspace{-6.9cm}

\noindent
{\footnotesize {\bf Figure SUR1-1.} Light curve of C/2012 S1, as of October
13, 2013, plotted against time, in comparison with the light curve of
C/1962~C1 (upper panel); and plotted against heliocentric distance, in
comparison with the light curves of C/1962~C1 and C/2002~O4 (lower
panel).}{\vspace{2.1cm}\pagebreak}
\end{figure}

Since the light curve of C/1962 C1 in the lower panel of Fig.~SUR1-1 shows
that between about 1.5 and 1.2~AU from the Sun the comet's intrinsic total
brightness varied with heliocentric distance $r$ approximately as $r^{-8}$,
it is tempting to estimate the power $n$ of heliocentric distance with
which does the brightness of C/2012~S1 vary at present.  This is easily
found out from the day-to-day variations in the normalized magnitude
$H_\Delta$.  In a parabolic approximation, the expression for $n$ in
the brightness law $r^{-n}$ is related to the daily rate of intrinsic
brightening (or fading), $dH_\Delta/dt$, by
\begin{equation}
n = \mp \, \frac{\sqrt{2}}{5 k \log e} \, r^{\frac{3}{2}} \!\! \left( 1 \! -
 \! \frac{q}{r} \right)^{\!\!-\!\frac{1}{2}} \! \frac{dH_\Delta}{dt} \simeq
 \mp \, 37.86 \, r^{\frac{3}{2}} \frac{dH_\Delta}{dt},
\end{equation}
where the minus sign applies before perihelion and the plus sign after
{\vspace{-0.05cm}}perihelion, $k$ is the Gaussian gravitational constant,
0.0172021~AU$^{\frac{3}{2}}\!$/day, \mbox{$\log e = 0.434294\,$\ldots},
and $q$ is the comet's perihelion distance in AU.  The simplified
expression on the right of (1) offers a satisfactory approximation as
long as \mbox{$r \gg q$}.% {\vspace{9cm}\pagebreak}

\begin{table}[h]
\noindent
\vspace{-0.1cm}
\begin{center}
{\footnotesize {\bf Table SUR1-1}\\[0.08cm]
{\sc Original Reciprocal Semimajor Axis for Comet C/2012~S1\\As Function of
 Orbital Solution's End Date.}\\[0.10cm]
\begin{tabular}{l@{\hspace{0.08cm}}c@{\hspace{0.02cm}}r@{\hspace{0.33cm}}c@{\hspace{-0.04cm}}c@{\hspace{0.04cm}}l}
\hline\hline\\[-0.22cm]
\multicolumn{3}{@{\hspace{0cm}}c}{End date} & Original reciprocal
 & Number & \\[-0.03cm]
\multicolumn{3}{@{\hspace{0cm}}c}{of orbital} & semimajor axis,
 & of observa- & \\[-0.02cm]
\multicolumn{3}{@{\hspace{0cm}}c}{solution}
 & $(1/a)_{\rm orig}$\,(AU$^{-1}$) & tions used
 & $\;$Reference$^{\rm a}$\\[0.05cm]
\hline\\[-0.22cm]
2013 & Aug.  & 23 & $+0.000\:028\,6\pm0.000\:001\,4$ & 3746 & E\,2013-Q27 \\
     & Sept. & 6 & $+0.000\:009\,2\pm0.000\:001\,2$ & 3897 & E\,2013-R59 \\
     &       & 16 & $+0.000\:008\,6\pm0.000\:001\,1$ & 3997 & MPC\,84932 \\
     &       & 30 & $+0.000\:005\,4\pm0.000\:000\,7$ & 4308 & E\,2013-S75 \\
     & Oct.  & 14 & $+0.000\:003\,8\pm0.000\:000\,6$ & 4677
             & E\,2013-T11\rlap{0} \\[0.09cm]
\hline\\[-0.22cm]
\multicolumn{6}{l}{\parbox{8.26cm}{$^{\rm a}$\,\scriptsize MPC = Minor Planet
 Circular; E = Minor Planet Electronic Circular (MPEC).}}\\[-0.15cm]
\end{tabular}}
\end{center}
\end{table}

An average rate of intrinsic brightening between October~4 and 13, at
heliocentric distances 1.59 to 1.40~AU, amounted to \mbox{$-0.030 \pm
0.010$}~mag/day, which, with an average heliocentric distance of 1.50~AU
in this period of time, is equivalent to \mbox{$n \simeq 2.1 \pm 0.7$}, a
considerably slower rate of upswing than was displayed by C/1962~C1.  At
this rate of brightening, C/2012~S1 would be appreciably fainter intrinsically
than C/1962~C1 near the Sun.  However, this trend may not necessarily
continue.  And even if it does, the smaller perihelion distance of C/2012~S1
and its somewhat stronger forward-scattering effect near perihelion, with the
phase angle peaking at 128$^\circ$ around December~1, should help offset the
shortfall.  The big question is:\ To what extent?

A new orbital solution that includes published astrometric observations up
to October 14, has been added to the previous ones in Table SUR1-1.  The
new solution shows that the negative trend in the original reciprocal
semimajor axis continues but is not accelerating.  Starting with the end
dates in early September, {\vspace{-0.03cm}}the average rate amounts to about
\mbox{$-4 \!\times\!  10^{-6}$}~AU$^{-1}$ per month, some two orders of
magnitude below that for C/2002~O4.

To appraise the comet's current status, I would suggest that, overall,
C/2012~S1 is looking reasonably healthy but not exuberant.}
\clearpage
\mbox{ }\\[-0.8cm]
\begin{center}
{\large \bf STATUS UPDATE REPORT \#2}\\
{\large \bf (October 22, 2013)}\\
\end{center}
{\normalsize
This status update report covers the period of time from October~13,
the end date of report \#1, to October~21.  Very few new total magnitude
observations of C/2012~S1 were published.  The count of the data points
used in the upgraded light curve, both versions of which are plotted in
Figure~SUR2-1, is now 49, based on reports from seven observers.

\begin{figure}[b]
\vspace{-1.15cm}
\hspace{-0.27cm}
\centerline{
\scalebox{0.475}{
\includegraphics{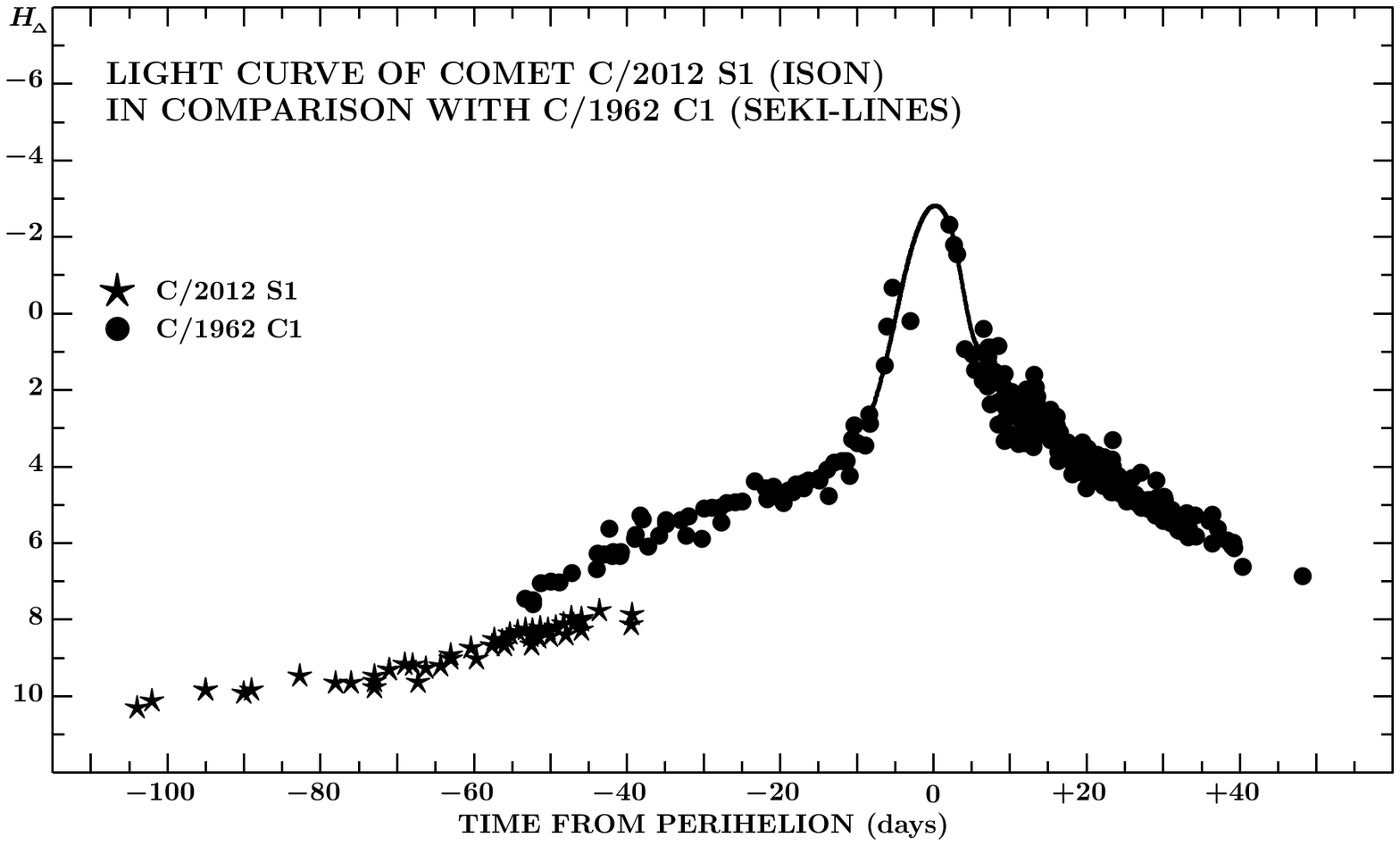}}} % from net_2012S1.tex

\vspace{-8.3cm}
\hspace{-0.31cm}
\centerline{
\scalebox{0.475}{
\includegraphics{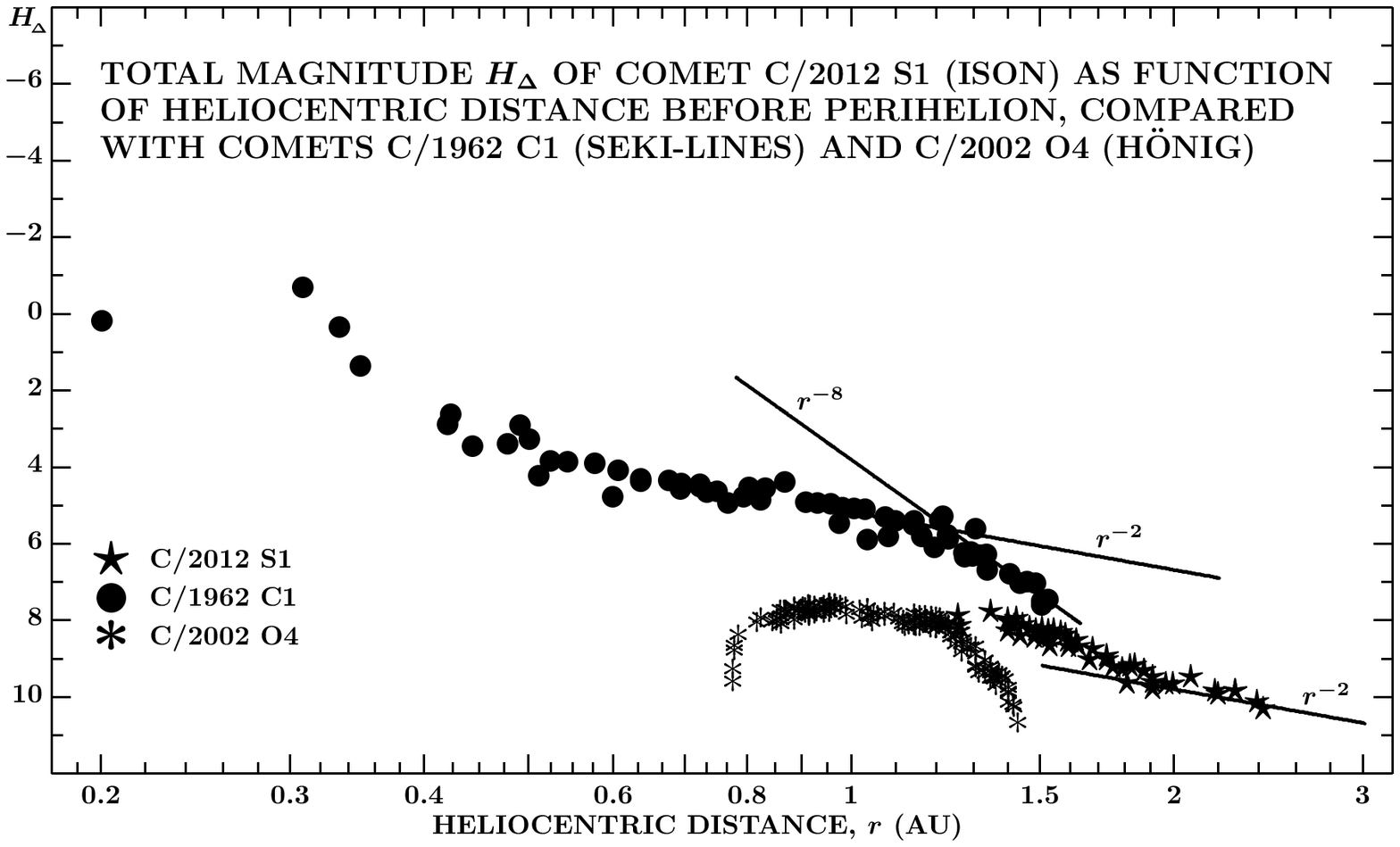}}} % from net_2012S1r.tex

\vspace{-6.9cm}

\noindent
{\footnotesize {\bf Figure SUR2-1.} Light curve of C/2012 S1, as of October
21, 2013, plotted against time, in comparison with the light curve of
C/1962~C1 (upper panel); and plotted against heliocentric distance, in
comparison with the light curves of C/1962~C1 and C/2002~O4 (lower
panel).}{\vspace{4.64cm}{\pagebreak}}
\end{figure}

The rate at which the comet's intrinsic brightness has lately been increasing
is fittingly described as a crawl.  An average daily rate of brightening
over the period of October~8--21, equals \mbox{$dH_{\!\Delta}/dt = -0.036
\pm 0.012$}~mag/day, which, with the mean heliocentric distance of 1.39~AU,
gives for an equivalent average power of intrinsic brightness variation with
heliocentric distance \mbox{$n = 2.2 \pm 0.7$}, practically the same as
before (see {\it SUR\,\#1\/}).  This means that the comet's activity generates
about as much gas (that radiates in the visible spectral region) and dust
as is lost, respectively, by photodissociation and escape of dust particles
into the tail.  Figure~SUR2-1 shows that at a heliocentric distance of
1.26~AU, the comet is about 2~magnitudes fainter intrinsically than comet
C/1962~C1 and only marginally brighter than comet C/2002~O4.  The latter
comparison should not, however, be interpreted to indicate that C/2012~S1 is
about to fizzle, as its light curve shows no clear signs of having peaked.  On
the other hand, a future outburst excepting, the comet's continuing lethargic
brightening does not make the prospects for its spectacular appearance near
the Sun any likelier than they were a week ago.

\begin{table}[h]
\noindent
\vspace{-0.1cm}
\begin{center}
{\footnotesize {\bf Table SUR2-1}\\[0.08cm]
{\sc Original Reciprocal Semimajor Axis for Comet C/2012~S1\\As Function of
 Orbital Solution's End Date.}\\[0.10cm]
\begin{tabular}{l@{\hspace{0.04cm}}c@{\hspace{0.02cm}}r@{\hspace{0.3cm}}c@{\hspace{-0.04cm}}c@{\hspace{0.04cm}}l}
\hline\hline\\[-0.22cm]
\multicolumn{3}{@{\hspace{0cm}}c}{End date} & Original reciprocal
 & Number & \\[-0.03cm]
\multicolumn{3}{@{\hspace{0cm}}c}{of orbital} & semimajor axis,
 & of observa- & \\[-0.02cm]
\multicolumn{3}{@{\hspace{0cm}}c}{solution}
 & $(1/a)_{\rm orig}$\,(AU$^{-1}$) & tions used
 & $\;$Reference$^{\rm a}$\\[0.05cm]
\hline\\[-0.22cm]
2013 & Aug.  & 23 & $+0.000\:028\,6\pm0.000\:001\,4$ & 3746 & E\,2013-Q27 \\
     & Sept. & 6 & $+0.000\:009\,2\pm0.000\:001\,2$ & 3897 & E\,2013-R59 \\
     &       & 16 & $+0.000\:008\,6\pm0.000\:001\,1$ & 3997 & MPC\,84932 \\
     &       & 30 & $+0.000\:005\,4\pm0.000\:000\,7$ & 4308 & E\,2013-S75 \\
     & Oct.  & 14 & $+0.000\:003\,8\pm0.000\:000\,6$ & 4677
             & E\,2013-T11\rlap{0} \\
     &       & 15 & $+0.000\:003\,6\pm0.000\:000\,5$ & 4688
             & MPC\,85336 \\
     &       & 21 & $+0.000\:004\,6\pm0.000\:000\,5$ & 4789
             & E\,2013-U17 \\[0.09cm]
\hline\\[-0.22cm]
\multicolumn{6}{l}{\parbox{8.26cm}{$^{\rm a}$\,\scriptsize MPC = Minor Planet
 Circular; E = Minor Planet Electronic Circular (MPEC).}}\\[-0.1cm]
\end{tabular}}
\end{center}
\end{table}

Two new orbital solutions have become available since {\it SUR\,\#1\/}.
Together with previous solutions, whose start date was September~30, 2011,
they are in Table SUR2-1.  The most recent one is based on published
astrometric observations up to October~21.  The new data for the
original reciprocal semimajor axis from the four most recent runs show
that the trend toward smaller values has essentially stopped, which means
that the sublimation-driven nongravitational perturbations of the comet's
orbital motion have not in the past three weeks increased with time,
perhaps due in part to the low activity.

The trend in the evolution of activity of C/2012~S1 during the past week
tends to reinforce the author's statement about the comet's health, as
expressed in {\it SUR\,\#1\/}.  The comet continues to brighten at a
sluggish rate, but has not run out of breath.
\clearpage
\mbox{ }\\[-0.97cm]
\begin{center}
{\large \bf STATUS UPDATE REPORT \#3}\\
{\large \bf (November 4, 2013)}\\
\end{center}
{\normalsize
This status update report covers the period of time from October~21,
the end date of report \#2, to November 2.  The number of observations
has in this period been steadily increasing, with new developments in
the comet's activity now apparent.

The count of the total-magnitude determinations used in the upgraded
light curve, both versions of which are plotted in Figure~SUR3-1, is now
92, based on reports from thirteen observers.  Comet C/2012~S1 is currently
about 3~magnitudes intrinsically fainter than was C/1962~C1 at the same
heliocentric distance.  Over the past week or so, the light curve of C/2012~S1
has nearly coincided with that of C/2002~O4.  In order to avoid an overlap,
the individual data points for C/2002~O4 were replaced with their mean curve
in the lower panel.  The coincidence should not be interpreted to indicate
that C/2012~S1 is near collapse.  It even cannot be ruled out that the
comet might partly recover its activity.

\begin{figure}[b]
\vspace{-2cm}
\hspace{-0.27cm}
\centerline{
\scalebox{0.475}{
\includegraphics{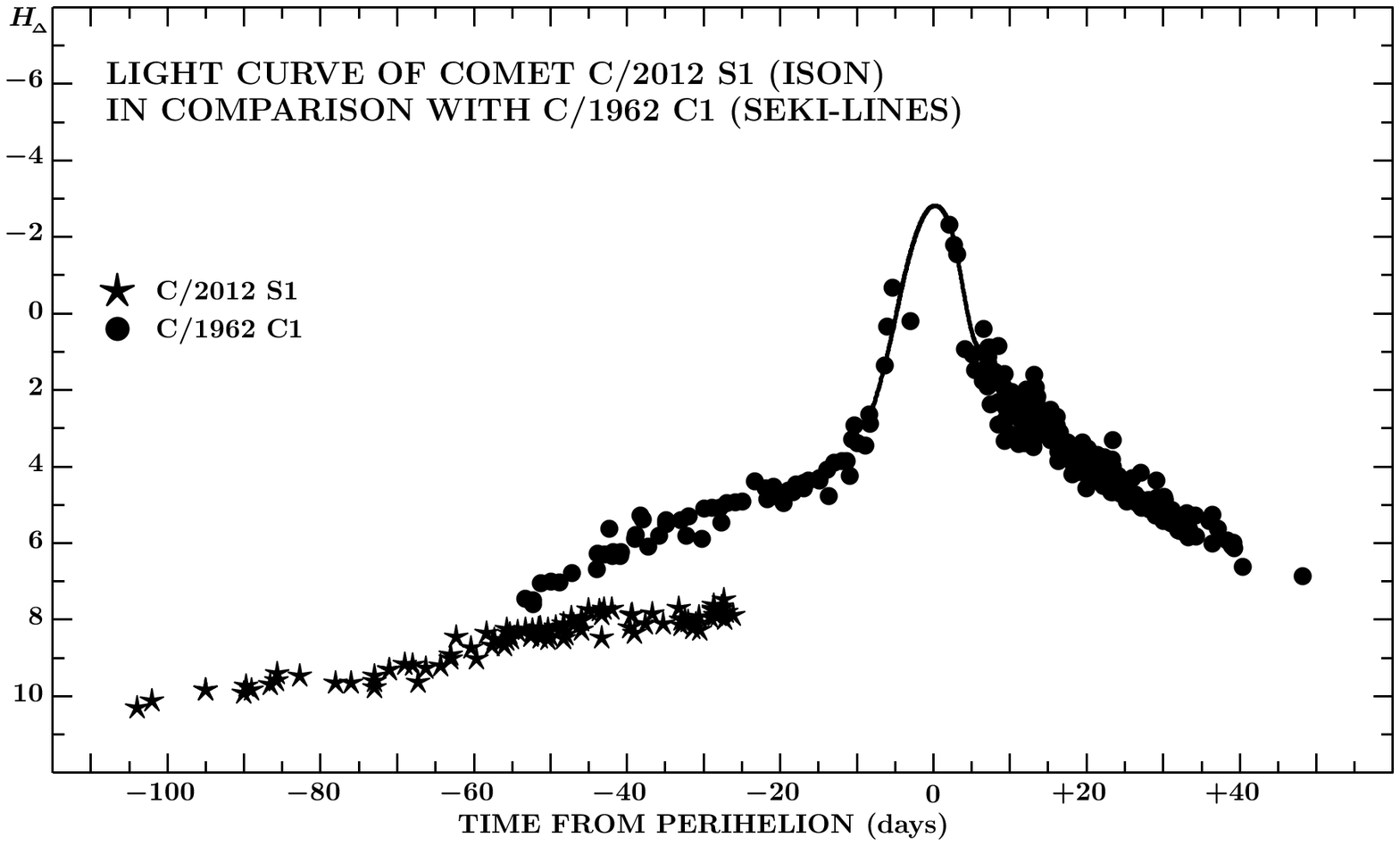}}} % from net_2012S1.tex

\vspace{-8.3cm}
\hspace{-0.31cm}
\centerline{
\scalebox{0.475}{
\includegraphics{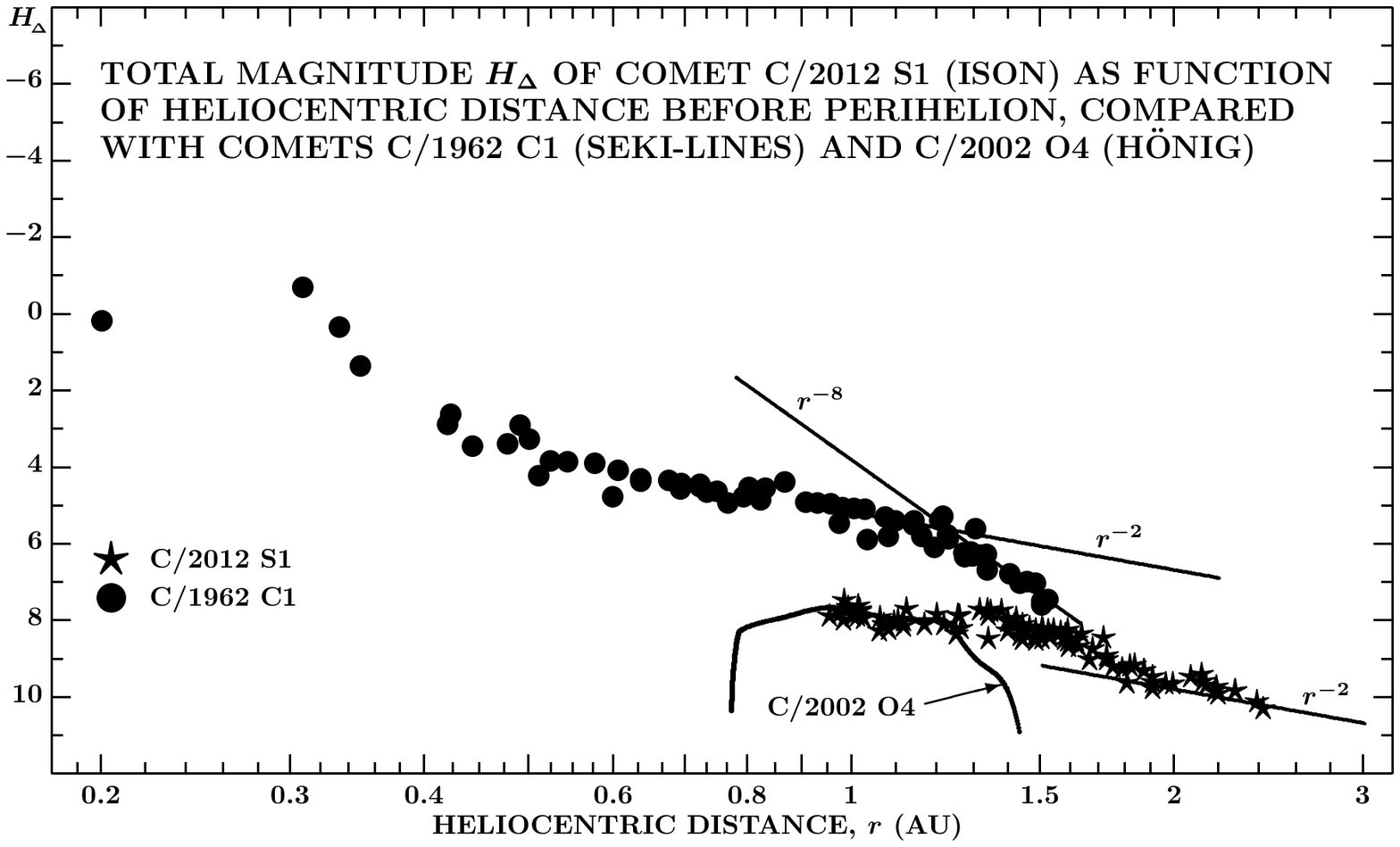}}} % from net_2012S1rr.tex

\vspace{-6.85cm}

\noindent
{\footnotesize {\bf Figure SUR3-1.} Light curve of C/2012 S1, as of November
2, 2013, plotted against time, in comparison with the light curve of
C/1962~C1 (upper panel); and plotted against heliocentric distance, in
comparison with the light curves of C/1962~C1 and C/2002~O4 (lower panel).}
\end{figure}

It appears that the comet's intrinsic brightness has nearly stagnated ever
since $\sim$October 13.  An average daily rate of brightening over the
period of October~20--November~2 equals \mbox{$dH_{\!\Delta}/dt = -0.023
\pm 0.009$}~mag/day, equivalent at an average of 1.11~AU from the Sun to
a~power of heliocentric distance of \mbox{$n = 1.0 \pm 0.4$},
which{\vspace{0.2cm}\pagebreak} means
that the total cross-sectional area of ejecta in the coma has recently been
declining and the comet has failed to fully resupply it and compensate for
the losses.

This conclusion is fundamentally consistent with the results from water
production measurements made on seven occasions between September~14 and
October~28 (D.\,Schleicher, IAUC 9260; H.\,Weaver et al., CBET 3680;
J.\,V.\,Keane et al., IAUC 9261; M.\,J.\,Mumma et al., IAUC 9261; and
N.\,Dello Russo et al., CBET 3686) and with the results from $Af\rho$
measurements made on five occasions between October 5 and 27 (A.\,Fitzsimmons
and P.\,Lacerda, IAUC 9261).

To the extent that the data points resulting from three different methods
of measuring H$_2$O in the coma can be combined, they show that the water
production rate has been stalling at 10$^{28.17 \pm 0.13}$\,molecules/s
over the more than six-week period.  Assuming that the sublimation rate
at a given heliocentric distance depends only on the Sun's zenith angle
as seen from the nucleus, an integration of the modeled sublimation rate
over the entire sunlit hemisphere suggests that in mid-September, at
1.95~AU from the Sun, the total water production area was 9.5~km$^2$,
while by late October, at 1.08~AU from the Sun, it was reduced to
2.4~km$^2$.  The equivalent diameters are, respectively, 2.46 and 1.24~km.

Fitzsimmons and Lacerda find from the quantity $Af\rho$ that over the
three-week period, when the comet was between 1.57 and 1.11~AU from the
Sun, the amount of dust in the coma was increasing as $r^{-0.3}$.

\begin{table}[h]
\noindent
\vspace{-0.2cm}
\begin{center}
{\footnotesize {\bf Table SUR3-1}\\[0.08cm]
{\sc Original Reciprocal Semimajor Axis for Comet C/2012~S1\\As Function of
 Orbital Solution's End Date.}\\[0.10cm]
\begin{tabular}{l@{\hspace{0.04cm}}c@{\hspace{0.02cm}}r@{\hspace{0.3cm}}c@{\hspace{-0.04cm}}c@{\hspace{0.04cm}}l}
\hline\hline\\[-0.22cm]
\multicolumn{3}{@{\hspace{0cm}}c}{End date} & Original reciprocal
 & Number & \\[-0.03cm]
\multicolumn{3}{@{\hspace{0cm}}c}{of orbital} & semimajor axis,
 & of observa- & \\[-0.02cm]
\multicolumn{3}{@{\hspace{0cm}}c}{solution}
 & $(1/a)_{\rm orig}$\,(AU$^{-1}$) & tions used
 & $\;$Reference$^{\rm a}$\\[0.05cm]
\hline\\[-0.22cm]
2013 & Aug.  & 23 & $+0.000\:028\,6\pm0.000\:001\,4$ & 3746 & E\,2013-Q27 \\
     & Sept. & 6 & $+0.000\:009\,2\pm0.000\:001\,2$ & 3897 & E\,2013-R59 \\
     &       & 16 & $+0.000\:008\,6\pm0.000\:001\,1$ & 3997 & MPC\,84932 \\
     &       & 30 & $+0.000\:005\,4\pm0.000\:000\,7$ & 4308 & E\,2013-S75 \\
     & Oct.  & 14 & $+0.000\:003\,8\pm0.000\:000\,6$ & 4677
             & E\,2013-T11\rlap{0} \\
     &       & 15 & $+0.000\:003\,6\pm0.000\:000\,5$ & 4688
             & MPC\,85336 \\
     &       & 21 & $+0.000\:004\,6\pm0.000\:000\,5$ & 4789
             & E\,2013-U17 \\
     &       & 28 & $+0.000\:007\,0\pm0.000\:000\,5$ & 4978
             & E\,2013-U73 \\
     & Nov.  &  2 & $+0.000\:009\,2\pm0.000\:000\,5$ & 5194
             & E\,2013-V07 \\[0.09cm]
\hline\\[-0.22cm]
\multicolumn{6}{l}{\parbox{8.26cm}{$^{\rm a}$\,\scriptsize MPC = Minor Planet
 Circular; E = Minor Planet Electronic Circular (MPEC).}}\\[-0.19cm]
\end{tabular}}
\end{center}
\end{table}

In the light of discouraging news on the activity of C/2012~S1, it is of
particular interest to see whether the comet's nucleus has still been
holding against the perturbations of its gravitational motion.  Over
the period since October 21, two new orbital solutions became available
as seen from Table SUR3-1.  There was no need to introduce the
nongravitational terms into the equations of motion, and the original
reciprocal semimajor axis actually continued to increase very slightly.
suggesting that as of November~2, the nucleus was in good shape.

In summary, although the performance of C/2012 S1 is close to anemic, there
still is no hard evidence that the comet is about to fall apart.  However,
prospects of a spectacular show near perihelion appear now to be less likely
than ever before.
\clearpage
\mbox{ }\\[-0.9cm] % 0.85
\begin{center}
{\large \bf STATUS UPDATE REPORT \#4}\\
{\large \bf (November 12, 2013)}\\
\end{center}
{\normalsize
This status update report, the last of the series, covers the period of time
from November~2, the end date of report \#3, to November 10.  The count of
the total-magnitude determinations used for the light curve from the period
of time starting on August~16 is now 111, based on reports by fifteen
observers.

The updated light curve, both versions of which are plotted in Figure~SUR4-1,
shows a new, encouraging sign:\ after about two weeks of stagnation and more
than three weeks of near-stagnation, a few days before the end of October the
comet's intrinsic brightening resumed and has now been sustained at a modest
rate of \mbox{$\langle dH_\Delta/dt \rangle = -0.065 \pm 0.007$}~mag/day,
judging from 39 observations spanning two weeks between October~27 and
November~10, which transforms into an average variation of \mbox{$r^{-2.24
\pm 0.24}$}.  Also significantly, the comet's light curve now runs parallel
to, though nearly 3~magnitudes below, the light curve of C/1962~C1.  On the
other hand, C/2012~S1 is now, in terms of heliocentric distance, beyond the
point of disintegration of C/2002~O4. ({\bf See an alert below, dated
November~14.})

\begin{figure}[b]
\vspace{-1cm}
\hspace{-0.27cm}
\centerline{
\scalebox{0.475}{
\includegraphics{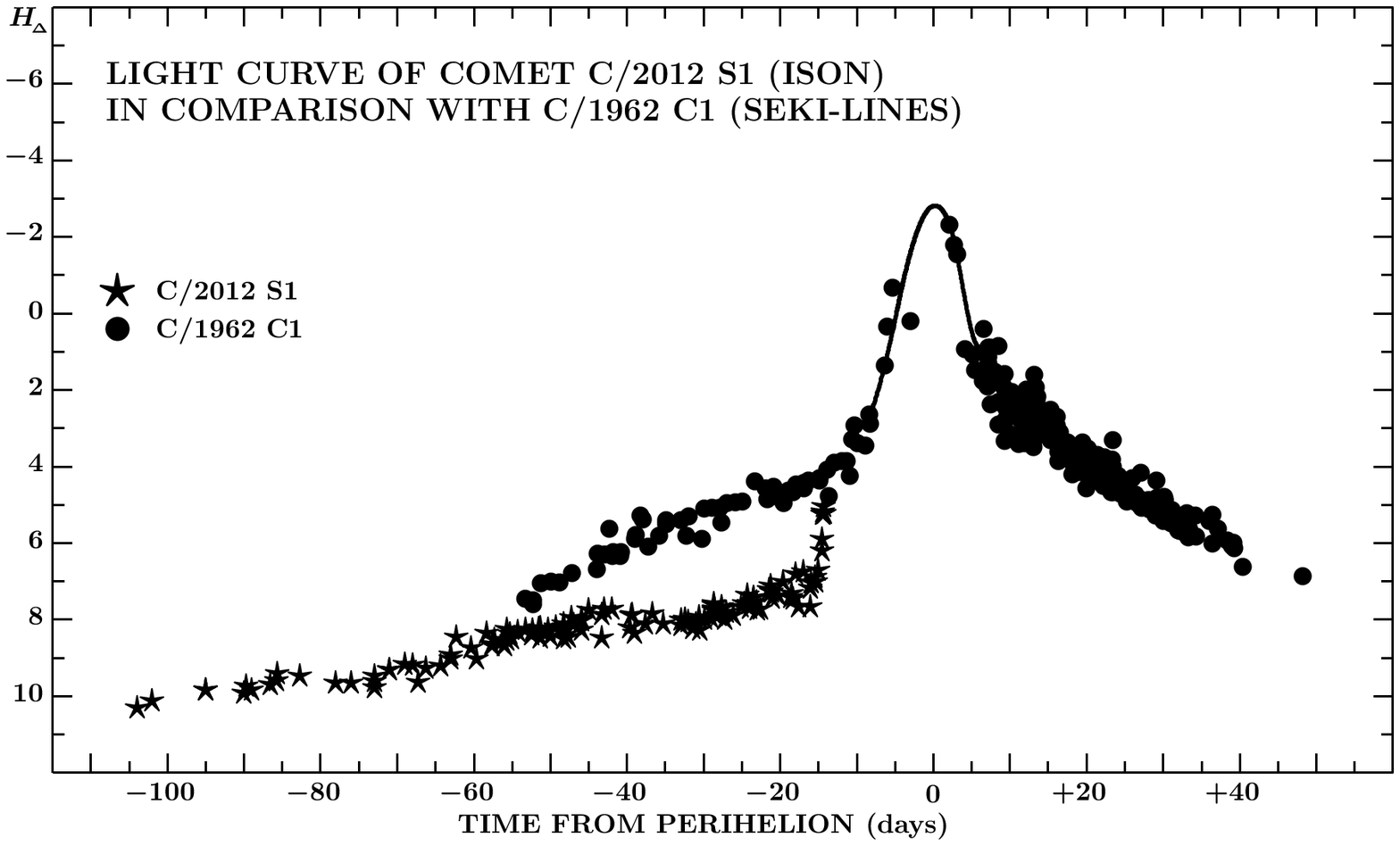}}} % from net_2012S1.tex

\vspace{-8.6cm} % 8.35
\hspace{-0.31cm}
\centerline{
\scalebox{0.475}{
\includegraphics{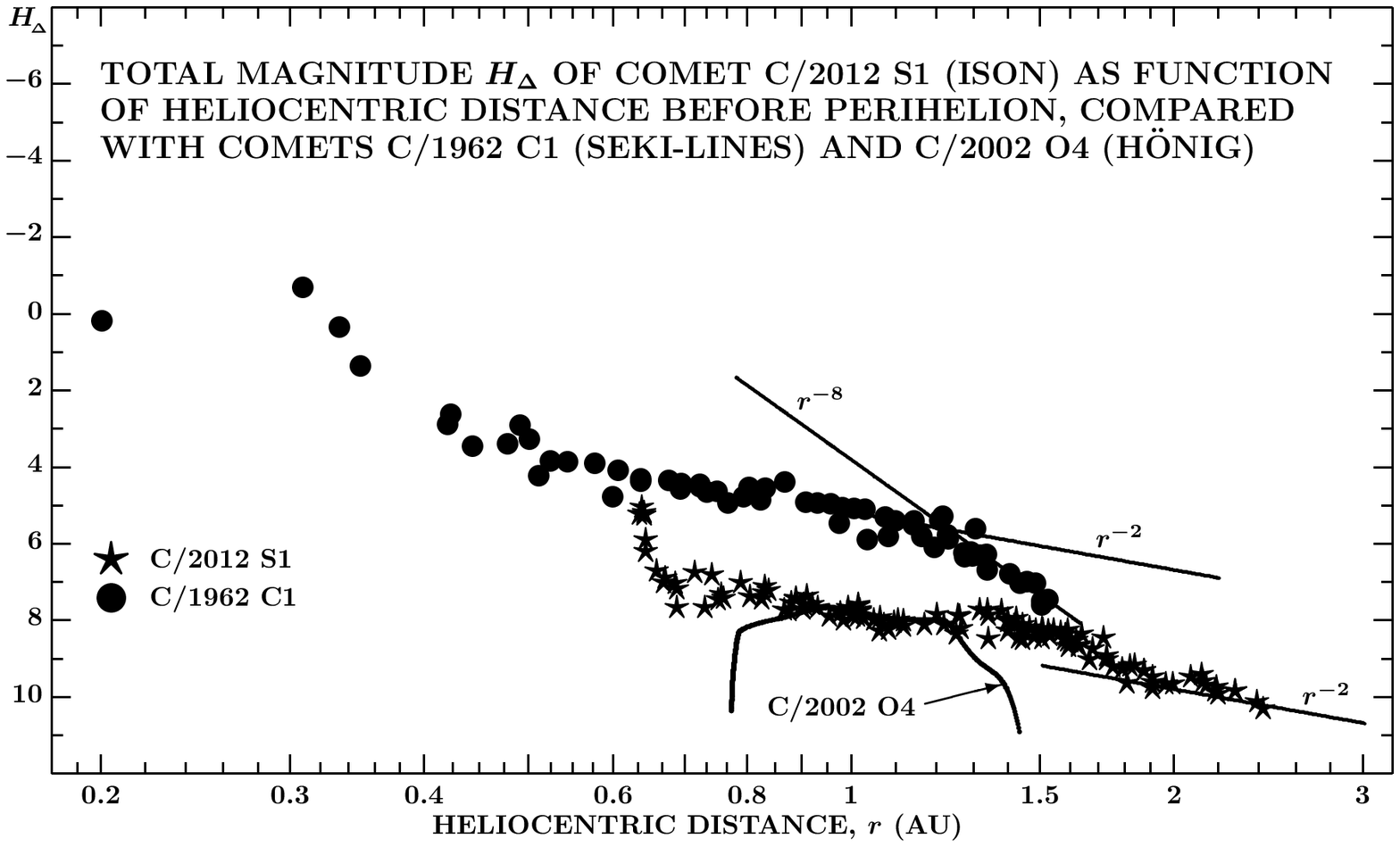}}} % from net_2012S1rr.tex

\vspace{-6.95cm} % 6.8

\noindent
{\footnotesize {\bf Figure SUR4-1.} Light curve of C/2012 S1, as of
November 10, 2013, plotted against time, in comparison with the light curve
of C/1962~C1 (upper panel); and plotted against heliocentric distance, in
comparison with the light curves of C/1962~C1 and C/2002~O4 (lower panel).}
({\bf Updated November 14.})
\end{figure}

The brightening is consistent with additional information on the activity
of C/2012~S1 that became available{\nopagebreak} in the past several days.
C.~Opitom et al.\ (CBET 3693) reported that the production of gas, such
as C$_2$ and CN, has increased rapidly since November~3.  Judging from
the OH emission, the water production grew as well, but because of large
error bars, this increase only slightly exceeded 1$\sigma$ between
October~31 and November~5.  No increase was detected in the production
rate of dust.

Very recently, H.~A.~Weaver et al.\ (web message)  reported a water
production rate of slightly exceeding \mbox{$2 \times 10^{28}$}\,molecules/s
derived from Hubble Space Telescope observations of the OH(0,0) band during
November~1, comparable to the rates they measured using the same technique
on October~8 and 21 (CBET 3680).  However, employing the Keck-2 telescope,
L.~Paganini et al.\ (IAUC 9263) determined from the near-infrared H$_2$O
emissions a water production rate of \mbox{$(3.1 \pm 0.2)
\times 10^{28}$}\,molecules/s on November~7, at least twice as high as on
October~22--25 (IAUC 9261).  This increase implies an $r^{-2.3 \pm 0.5}$
variation.

Some issues related to the comet's overall light curve, from 9.4~AU at the
end of September 2011 down to less than 0.8~AU before perihelion, are
briefly addressed in the Appendix to this paper, which follows.
% this last
% update.  The recently observed resumption of activity may very well be
% an integral part of a cycle of its variations that has recurred several
% times since the days of the 2011 prediscovery observations.

\begin{table}[h]
\noindent
\vspace{-0.3cm}
\begin{center}
{\footnotesize {\bf Table SUR4-1}\\[0.08cm]
{\sc Original Reciprocal Semimajor Axis for Comet C/2012~S1\\As Function of
 Orbital Solution's End Date.}\\[0.10cm]
\begin{tabular}{l@{\hspace{0.04cm}}c@{\hspace{0.02cm}}r@{\hspace{0.3cm}}c@{\hspace{-0.04cm}}c@{\hspace{0.04cm}}l}
\hline\hline\\[-0.22cm]
\multicolumn{3}{@{\hspace{0cm}}c}{End date} & Original reciprocal
 & Number & \\[-0.03cm]
\multicolumn{3}{@{\hspace{0cm}}c}{of orbital} & semimajor axis,
 & of observa- & \\[-0.02cm]
\multicolumn{3}{@{\hspace{0cm}}c}{solution}
 & $(1/a)_{\rm orig}$\,(AU$^{-1}$) & tions used
 & $\;$Reference$^{\rm a}$\\[0.05cm]
\hline\\[-0.22cm]
2013 & Aug.  & 23 & $+0.000\:028\,6\pm0.000\:001\,4$ & 3746 & E\,2013-Q27 \\
     & Sept. & 6 & $+0.000\:009\,2\pm0.000\:001\,2$ & 3897 & E\,2013-R59 \\
     &       & 16 & $+0.000\:008\,6\pm0.000\:001\,1$ & 3997 & MPC\,84932 \\
     &       & 30 & $+0.000\:005\,4\pm0.000\:000\,7$ & 4308 & E\,2013-S75 \\
     & Oct.  & 14 & $+0.000\:003\,8\pm0.000\:000\,6$ & 4677
             & E\,2013-T11\rlap{0} \\
     &       & 15 & $+0.000\:003\,6\pm0.000\:000\,5$ & 4688
             & MPC\,85336 \\
     &       & 21 & $+0.000\:004\,6\pm0.000\:000\,5$ & 4789
             & E\,2013-U17 \\
     &       & 28 & $+0.000\:007\,0\pm0.000\:000\,5$ & 4978
             & E\,2013-U73 \\
     & Nov.  &  2 & $+0.000\:009\,2\pm0.000\:000\,5$ & 5194
             & E\,2013-V07 \\
     &       &  8 & $+0.000\:009\,6\pm0.000\:000\,5$ & 5363
             & E\,2013-V48 \\[0.09cm]
\hline\\[-0.22cm]
\multicolumn{6}{l}{\parbox{8.26cm}{$^{\rm a}$\,\scriptsize MPC = Minor Planet
 Circular; E = Minor Planet Electronic Circular (MPEC).}}\\[-0.2cm]
\end{tabular}}
\end{center}
\end{table}

The resumed activity of comet C/2012 S1 appears to have had, as of November~8,
no effect on its orbital motion, as is shown by comparing the original
reciprocal semimajor axis from the most recent orbital solution with the
previous entries in Table SUR4-1.  The comet's motion still is satisfactorily
matched by a purely gravitational orbit.

To summarize, the immediate future of C/2012~S1 looks a little brighter
now than a week ago, but the prospects of an exceptionally striking display
near and after perihelion are still not good, unless the nucleus suddenly
disintegrates near or shortly after perihelion into a cloud of dust.  The
forward scattering effect should increase the brightness by up to
$\sim$1~magnitude around December~1, but very little (probably $<$0.2
magnitude) before perihelion.

{\bf Alert}: Starting at $\sim$5$^{\rm h}$ UT on November~14, comet C/2012~S1
has been reported to be in major outburst.  From early data, a preliminary
estimate for its onset~is November~14.$0 \pm 0$.2~UT, with an amplitude of
at least 2~mag.  Intrinsically the comet is now almost as bright~as
C/1962~C1 at the same heliocentric distance.  It is unclear whether this
event's nature is benign or cataclysmic.

\begin{table*}[ht]
\vspace{0.2cm}
\begin{center}
{\large \bf Appendix}\\[0.2cm]
{\large \bf LIGHT CURVE OF COMET C/2012 S1 (ISON) AS\\[-0.01cm]
A SEQUENCE OF SEVERAL CONSECUTIVE CYCLES OF\\[0.06cm]
TWO-STAGE ACTIVITY EVOLUTION}{\nopagebreak}
\end{center}
\end{table*}
In the paper and all the updates I was concerned only with the light
curve since the reappearance of comet C/2012~S1 after the July 2013
conjunction with the Sun, because the main issues were the comet's
comparison with the other two objects and its changing behavior on
relatively short time scales.  In this Appendix I present the
comet's light curve in its entirety, starting with the prediscovery
observations at the end of September 2011.\footnote{See {\tt
http://www.minorplanetcenter.net/db\_search}.}  The sources of data were
the web site of the {\it International Comet Quarterly\/},\footnote{See
{\tt http://www.icq.eps.harvard.edu/CometMags.html}.} the web site of
the {\it Comet Observations Group\/},\footnote{See {\tt
http://groups.yahoo.com/neo/groups/CometObs/info}.} and some of the CCD
sets of {\it total\/} magnitudes (T) reported to the {\it Minor Planet
Center\/}.$^2$  The procedure that established a common photometric
system was already described in the paper itself (Sec.~2).

The assembled light curve, presented in the following --- just as in the
paper and the updates --- both as a plot against time (reckoned from the
perihelion time) and as a plot against heliocentric distance, is based on
227~total-magnitude determinations\footnote{Sets of brightness data points
from the same day by the same observer(s) have been consistently averaged
into a~single data point and are counted and plotted as such.  The only
exceptions are the rather discordant prediscovery magnitudes reported
from the November 26 and December 9, 2011 observations by the Pan-STARRS~1
Station at Haleakala; all four data points, two on each date, were
averaged into a~single data point.} by 29~observers (or groups of
observers).

As a dynamically new comet, C/2012 S1 is believed to have originated, with
countless others, by accretion of a~primordial material of the protoplanetary
nebula in regions near Jupiter's orbit and was afterward ejected by
perturbations into the Oort Cloud, in which, over billions of years, was in
an extremely cold environment mercilessly bombarded by galactic cosmic rays.
A~relatively thin outer layer of the nucleus was saturated with free
radicals and other chemically active species.  During a~journey to
the inner solar system, even slight warming of the surface of comets
like C/2012~S1 at very large heliocentric distances leads to release of
reactive species from the surface layer, making these objects unusually
active (compared to dynamically old comets).  For example, their
apparently icy tails made up of submillimeter- to millimeter-sized grains
are formed at heliocentric distances of up to at least 15~AU, as is readily
inferred from the tails' strong deviation from the prolonged radius vector
for objects with perihelia beyond 2--3~AU.

Hyperactivity of dynamically new comets results, however, in their highly
volatile species being rapidly exhausted, which in turn brings about
a~decline in the ejection rate of material from the nucleus and sometimes
even a temporary drop in the comet's brightness during the continuing
approach to the Sun.  The light curve of comet C/2012~S1 is a~superb
illustration of the complexity of a new comet's activity and physical
behavior.

Figure A-1 shows the variations in the comet's total intrinsic brightness
as a function of time.  It is immediately evident that the light curve
consists of at least four (and possibly five) consecutive periods, in each
of which the brightness first increases, reaches a~local maximum before
stagnating or subsiding.  This means that the comet's activity evolves
literally in cycles, each of which begins with an {\it ignition point\/},
introducing an {\it active stage\/}, and terminates with a~{\it stage
of progressive deactivation\/}.  In the first cycle (A), the activity is
controled by the sublimation of the most volatile ice available in abundance
and continues until its supplies are depleted.  From that time on, the
activity is governed by the sublimation of the next ice in a~succession of
diminishing volatility, etc., until eventually the least volatile, water ice
takes its turn.  Each cycle requires a new source.

Even though the light-curve data are very incomplete before the discovery
of C/2012~S1, it is possible to estimate the brightness variations in the
major gap between 430 and 670 days before perihelion thanks to insignificant
changes  after the 240~days.  A~similar development took apparently place
during the comet's conjunction with the Sun between mid-June and mid-August
2013 (110--160 days before perihelion).  Besides, the comet's behavior
throughout a~stage of progressive deactivation is readily perceived from the
observations in cycle B between mid-January and the beginning of May 2013.

The properties of a cycle are suggested by the ignition-point position.
The beginning of cycle~A is unknown, but it must
have occurred more than 790~days before perihelion, more than 9.4~AU from
the Sun.  The ignition point of cycle~B cannot be determined accurately,
but it took place most probably between 460 and 550~days before perihelion,
6.5 to 7.4~AU from the Sun.  Cycle~C began around 212~days before perihelion
and 3.9~AU from the Sun, and cycle~D some 115~days before perihelion and
2.6~AU from the Sun.  It is well-known that the sublimation of water ice
usually begins to dominate comet activity at heliocentric distances between
2 and 3~AU, so that cycle~D in Figures~A-1 and A-2 is very probably water-ice
controled.  In a so-called isothermal water sublimation model an ice-covered
surface of the nucleus has a~temperature of 170~K at a distance of 2.6~AU
from the Sun.  The ignition points in cycles~A and B are deemed to refer
to activity governed by highly volatile ices, such as carbon monoxide and
carbon dioxide.  In reality, the situation is more complex.  In an extreme
case, with the Sun constantly near the zenith, the water sublimation can
control activity even near 5~AU from the Sun.

Figure A-2 shows that during the active stage of each cycle (with a~possible 
exception of the less confidently identified cycle~E) the brightness grew
with a fairly high power of heliocentric distance, between $\sim$4 and
$\sim$6.  This brightening was however largely mitigated by a~stagnation
or drop of activity during each deactivation stage, so that overall the
intrinsic brightness grew only as $\sim \! r^{-2}$.
\clearpage
\begin{figure*}
\vspace{-3.6cm}
\hspace{-0.4cm}
\centerline{
\scalebox{0.867}{
\includegraphics{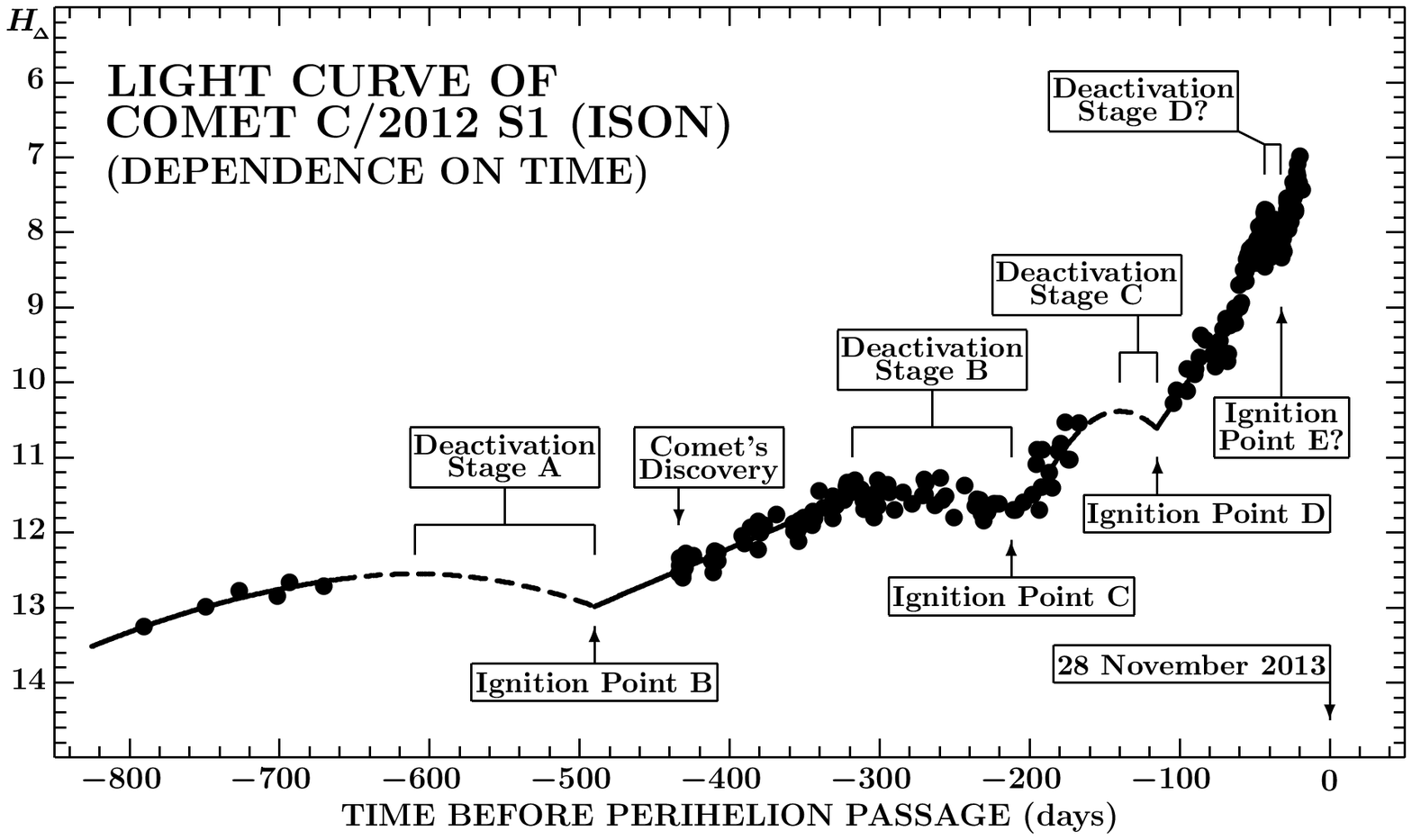}}} % from fA1_2012S1.tex
\vspace{-12.15cm}

{\footnotesize {\bf Figure A-1}.  The entire light curve of C/2012~S1, as
it appeared as of November 10, 2013, 18~days before perihelion, plotted as a
function of time, reckoned from the time of perihelion passage.  Temporal
variations in the normalized total brightness, expressed by the magnitude
$H_\Delta$, consist of at least four, and possibly five, consecutive
sections --- quasi-periodic cycles A, B, C, D, and apparently also E ---
each of which begins with an ignition point, introducing an active
stage, and terminates with a stagnation or drop in a stage of progressive
deactivation.  Each cycle requires a new source of activity.}

\vspace{-2.6cm}
\hspace{-0.38cm}
\centerline{
\scalebox{0.87}{
\includegraphics{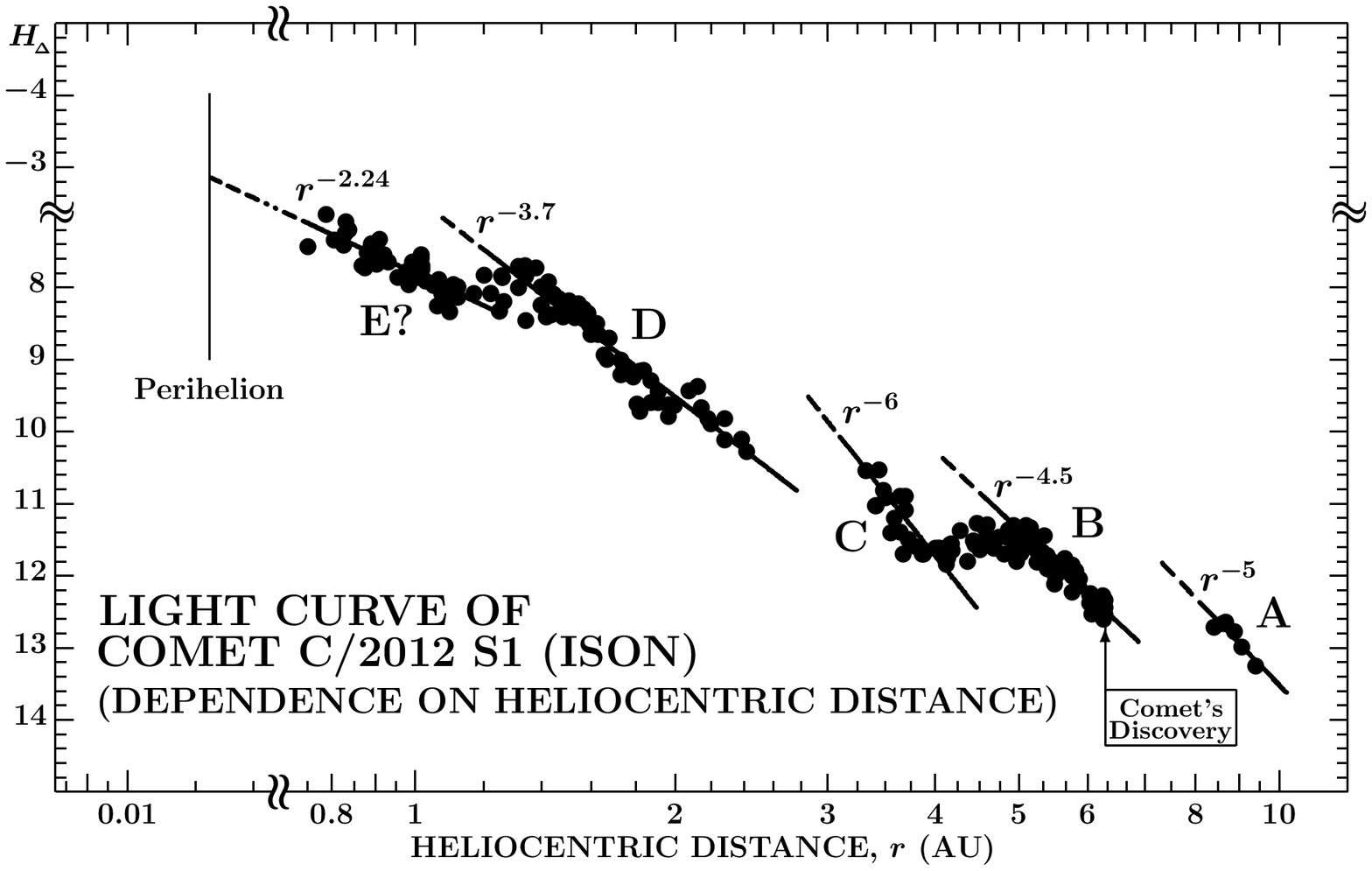}}} % from fA2_2012S1.tex
\vspace{-12.25cm}

{\footnotesize {\bf Figure A-2}.  The same light curve of C/2012~S1 plotted
as a function{\vspace{-0.06cm}} of heliocentric distance.  This relationship
is approximated in the active stages of the cycles A, B, C, D, as well as
E by an inverse dependence on a power $n$ of the distance, $r^{-n}$, where
$n$ is seen to be confined to a range from 2.24 to 6.  Note a stagnation of
activity after the active stages B and D.  Also note that the periods of
stagnation (or drop) drag down the average rate of brightening with decreasing
heliocentric distance, which is close to $r^{-2}$.}
\end{figure*}
\end{document}